\PassOptionsToPackage{dvipsnames}{xcolor}
\documentclass[acmsmall]{acmart}\makeatletter                   %1
\def\mdseries@tt{m}             %1
\makeatother                    %1\usepackage[plain]{fancyref}
\usepackage[draft=true]{minted} %2
\usepackage{color}
\usepackage{hyperref}           %5
\hypersetup{
    colorlinks=true,
    linkcolor=blue,
    filecolor=red,      
    urlcolor=magenta,
    breaklinks=true,            %3
}
\usepackage{breakurl}           %3

\usepackage{graphicx}
\acmConference[CSCW]{November 9-13th 2024}{San Jose, Costa Rica}

\setcopyright{acmlicensed}
\acmJournal{PACMHCI}
\acmYear{2024} \acmVolume{8} \acmNumber{CSCW2} \acmArticle{416} \acmMonth{11}\acmDOI{10.1145/3686955}

\begin{document}

\title[Attention is All They Need]{Attention is All They Need: Exploring the Media Archaeology of the Computer Vision Research Paper}

\author{Samuel Goree}
\email{sgoree@stonehill.edu}
\affiliation{
    \institution{Stonehill College}
    \city{Easton}
    \state{MA}
    \country{USA}
}

\author{Gabriel Appleby}
\email{gabriel.appleby@tufts.edu}
\affiliation{
    \institution{Tufts University}
    \city{Medford}
    \state{MA}
    \country{USA}
}

\author{David Crandall}
\email{djcran@indiana.edu}
\affiliation{
    \institution{Indiana University}
    \city{Bloomington}
    \state{IN}
    \country{USA}
}

\author{Norman Makoto Su}
\orcid{0000-0001-8765-6579}
\email{normsu@ucsc.edu}
\affiliation{
    \institution{University of California, Santa Cruz}
    \city{Santa Cruz}
    \state{CA}
    \country{USA}
}

\renewcommand{\shortauthors}{Goree, Appleby, Crandall, and Su}

\def\normsu#1{\textcolor{ForestGreen}{norm: #1}}
\def\djc#1{\textcolor{Blue}{djc: #1}}

\begin{abstract}
  Research papers, in addition to textual documents, are a designed interface through which researchers communicate.
  Recently, rapid growth has transformed that interface in many fields of computing.
  In this work, we examine the effects of this growth from a media archaeology perspective, through the changes to figures and tables in research papers.
  Specifically, we study these changes in computer vision over the past decade, as the deep learning revolution has driven unprecedented growth in the discipline.
  We ground our investigation through interviews with veteran researchers spanning computer vision, graphics, and visualization.
  Our analysis focuses on the research attention economy: how research paper elements contribute towards advertising, measuring, and disseminating an increasingly commodified ``contribution.''
  Through this work, we seek to motivate future discussion surrounding the design of both the research paper itself as well as the larger sociotechnical research publishing system, including tools for finding, reading, and writing research papers.
\end{abstract}




\begin{teaserfigure}
  \centering
  \includegraphics[trim={0.5in 1.4in 0.5in 1in},clip,width=0.8\textwidth]{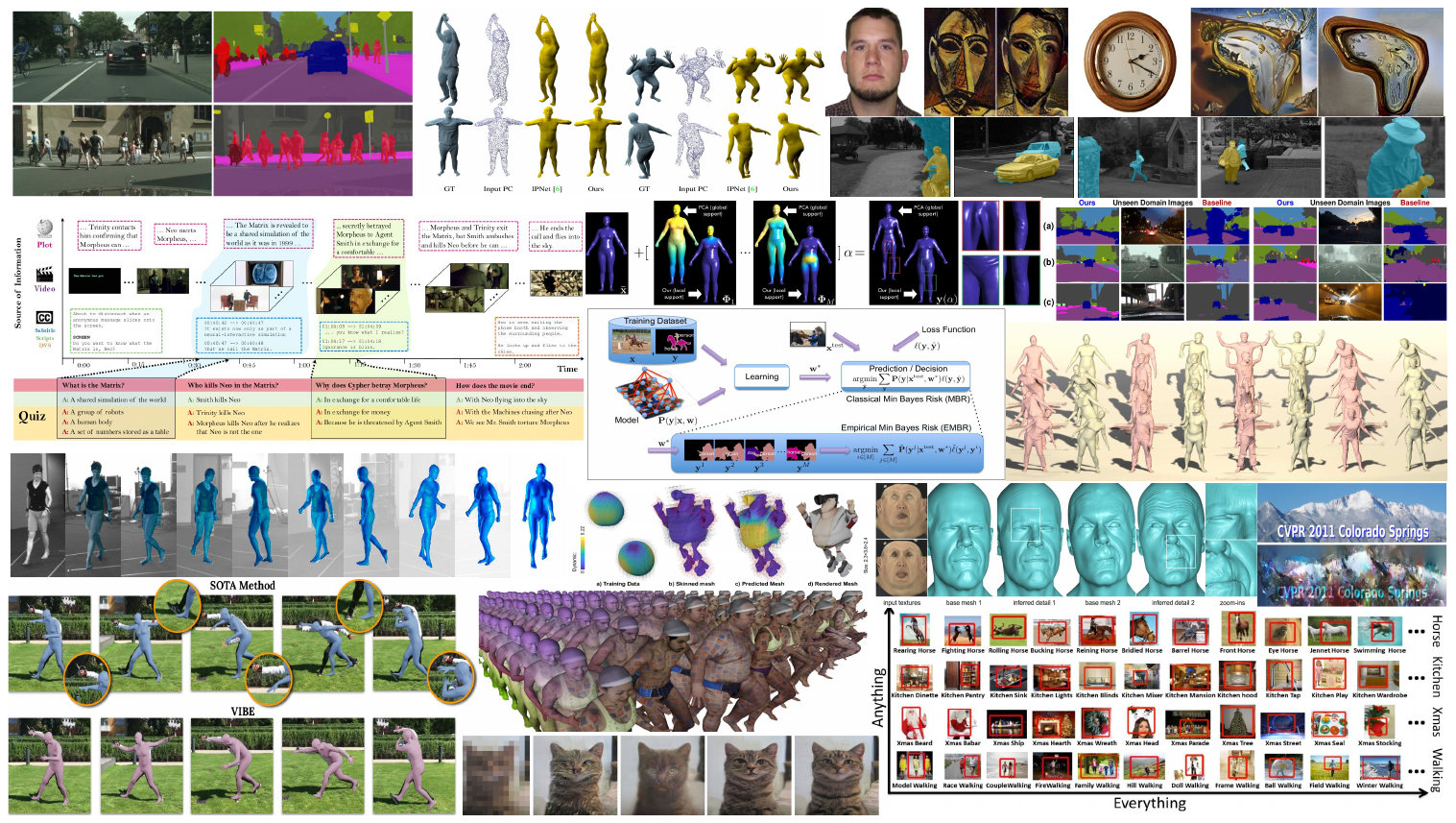}
  \caption{Teasers from computer vision papers \cite{4270338,5995616,6909814,6909533,7299055,7780976,7780714,7780870,8100074,8100065,8578975,9578478,9578823,9577906,9578521,9577713,9156519}. Best viewed in color.}
  \Description{Several teaser images from computer vision research papers, including elements like various 3D models of human faces and bodies, images with bounding boxes and distorted images.}
  \label{fig:teaser}
\end{teaserfigure}
\maketitle

\section{Introduction}

The research paper has many meanings. It is a contribution to
knowledge, often responding---according to formal and informal rules
of scientific discourse---to a decades-long conversation. It signals
to peers, funding agencies, future employers,
and future students that the authors are knowledgeable about a topic
and actively studying it. It is a unit of productivity which indicates
progress towards a PhD, tenure, or promotion. But beneath all of those
layers of interaction, it is a digital media artifact, an element of the
sociotechnical research publishing system, with a visual form that both 
influences, and is influenced by, its discipline.

We propose thinking about the research paper not as a
neutral medium for disseminating textual scientific content, but as a designed
media artifact with which researchers---its ``users''---interact with digitally. For
example, we might \textbf{glance} at its title in a list of papers, \textbf{click} on it,
get \textbf{hooked} by its first page figure,
\textbf{scroll} through its pages in a PDF viewer, \textbf{scanning} a table of results. We may also
\textbf{print} the paper and \textbf{carry} it with
us, \textbf{write} in its margin and leave it on a desk, and
then \textbf{find} it months or years later. In all of these
interactions, the design of both the visual and textual content, as
well as the design of the larger technical system, which produced the
PDF, delivered it to us, rendered its image to a screen and transferred it to paper, affects
and shapes our interactive attentive experience. Moreover, research papers are cultural objects shaped by disciplinary culture. Bruno Latour, for
example, writes extensively about the social nature of the research
paper and the important rhetorical role its figures and tables play in
creating scientific truth~\cite{latour1987science}. Taking this further, we
view the research paper as a media object
\cite{manovich2002language}, an assemblage of designed technologies---like rendered data visualizations, LaTeX, and
the PDF file format---viewed on a screen or printed on paper, which have technological affordances and, as we will discuss,
accessibility concerns.

In this paper, we study the changes to computing papers in one area---\textit{computer vision}---over the past decade. While changes occur over time in the research papers of every discipline, computer vision is a particularly information-rich site~\cite[p.\ 242]{Patton:1990} for understanding the paper as a media artifact both because of its
inherently visual nature and its unprecedented recent growth. Over the past decade, the ``deep learning revolution'' has transformed several fields of computer science altering both the
fields themselves and the way researchers and
practitioners feel about them~\cite{su2021affective}. In computer vision specifically, Su and Crandall find
``a general mood of malaise''
\cite{su2021affective} which permeates the field. We argue that this malaise is a symptom of increased commodification of scholarly attention, which
is materially documented in the changing design of research papers.

By attention, we evoke the attention economy, first proposed by Herbert Simon \cite{Simon1971} and applied to the internet by authors such as Goldhaber \cite{goldhaber1997attention} and
Davenport and Beck \cite{davenport2001attention}. An attention economy in the workplace has been discussed in CSCW, for example in the work of Yardi et al.\ on the attention economy of internal corporate blogs \cite{yardi2009blogging} or Kraut et al.\ on the attention economy of email marketing \cite{kraut2002markets}. To understand its commodification, we turn to the post-Marxist analysis of Claudio Celis Bueno \cite{buenoattention}. Bueno frames online attention not just as a scarce resource, but as a form of labor---human activity which has the potential to generate surplus value. We claim that this theory of attention as labor applies to the attention paid to academic research as well, and helps to explain recent changes to computer vision.

Research papers in computing have had a rich history of debate (e.g., conferences vs.\ journals~\cite{Vardi:2009}). In the field of HCI and its subfields, including CSCW, debates continue to arise regarding the form and content of the research paper~\cite{Geiger:2019,WallaceOjiEtAl:2017}. Canonical works like Dourish's ``Implications for Design'' comment on the relationship between scholarly publishing and larger issues of power in disciplinary culture \cite{dourish2006implications}. Sometimes, new formats arise in response; for instance, the pictorial format addresses the need to foreground visual imagery as a significant contribution to HCI~\cite{BlevisHauserEtAl:2015,Blevis:2016}. In HCI, these debates exist both as meta-scholarship, shaping the practices of a discipline, and as subject matter, studying the research publishing system as a technological platform.

Perhaps the biggest hint that the design of computer vision papers has changed with the rapid changes in its disciplinary culture is scholarly humor. In the 2010 computer vision satire paper ``Paper Gestalt''
\cite{von2010paper}, the pseudonymous authors ``take the simple
intuition that the quality of a paper can be estimated by merely
glancing through the general layout, and use this intuition to build a
system that employs basic computer vision techniques to predict if the
paper should be accepted or rejected'' \cite{von2010paper}, and suggest that this system
might replace the peer review process. The visual design of the paper
parodies the qualities it observes, including unnecessary complex
equations and long algorithms.

But in 2018, Jia Bin Huang published a sequel, ``Deep Paper Gestalt''
\cite{huang2018deep}, which updates both the methods and the jokes. Juxtaposing
these two papers gives a glimpse of how the style of computer vision
research papers has changed in just eight years. Instead of using overly complex algorithms and equations, Huang 
proposes a benchmark dataset, CVPG (the ``Computer Vision Paper Gestalt''
benchmark), and uses deep learning to significantly outperform the
``hand-crafted features'' of ``Paper Gestalt''~\cite{von2010paper}. Then, it
presents an unnecessarily dense table and gratuitously large
figure of class activation heatmaps to show the page regions which
predict good papers. Though satire, these papers raise real questions 
about the research paper as a media artifact.

%Though our topic is a bit unconventional, we believe that our findings
%are nevertheless highly relevant to HCI. First, we study researchers
%as users of electronic reading and writing systems, like PDF viewers,
%reference management tools, LaTeX, visualization libraries, and
%graphic design tools. A strong understanding of these users, their use
%patterns, and how they specifically change over time is important for
%the design of these systems. Second,  
%HCI researchers should understand how to read these papers in their
%subtly-different disciplinary context in order to use them effectively. 
%We believe that studying the design evolution of computer vision papers helps
%to build such understanding.

In the following sections, we examine historical computer vision research papers from a media archaeology perspective, focused on the way their visual
style has developed over time. To better understand these developments
in context, we also report results of interviews with twelve veteran
researchers whose work spans computer vision, graphics, and visualization. Our approach focuses on elements of the research paper and larger publication
system based on their roles in advertising, measuring, and disseminating an
increasingly commodified ``contribution.'' We find that as academic
discourse has moved online, the limiting factor on publishing has shifted
from printing costs to the attention of other researchers, which has
changed both the culture of computer vision as well as the design of its research papers. While this finding echoes existing discourse surrounding ``publish or perish'' \cite{garfield1996primordial,lee2014publish}, the ``marketization'' of academic discourse \cite{fairclough1993critical}, use of ``hype'' in science communication \cite{soto2023hype}, or promotional qualities in titles and abstracts \cite{hyland2023academic}, we emphasize the novelty of our approach. To the best of our knowledge ours is the first connection between these discussions of scholarly discourse and the designed visual artifact of the research paper.

\section{Related Work}

Our inquiry sits at the intersection of several scholarly conversations
in and around HCI and CSCW, 
including data mining of research paper figures,
studies of digital media objects, designing for readers and writers of
research papers, and studies of the culture of computer vision.

\subsection{Research papers as media objects}

The fascinating satire paper ``Paper Gestalt''~\cite{von2010paper} is
both a related work and a primary source in our inquiry. It claims,
facetiously, that ``the quality of a paper can be estimated by merely
glancing through the general layout'' and uses machine learning to
predict whether papers will be accepted or rejected from their page
images. The (unknown) authors identify that certain features, like
sophisticated math and large figures, are predictive of acceptance
while large, dense tables are predictive of rejection. A sequel,
Jia-Bin Huang's ``Deep Paper Gestalt'' \cite{huang2018deep}, updates
this methodology using deep learning. These papers are a commentary on the
the reality of the modern computer vision community: a reliance on visual
performance to ensure successful dissemination and influence of 
papers.

There is also work surrounding the analysis of research papers in
document image analysis~\cite{davila2020chart} and data
visualization~\cite{wu2021ai4vis,chen2021vis30k}. These methods,
however, treat the figure as a neutral data source, where there is an
objective correspondence between the original data and the figure for
each chart type. More recent work has started to, for example, see data visualization as not merely objective reporting of facts but as a communicative medium that has affective impact~\cite{Lee-RobbinsAdar:2022,vanKoningsbruggenHornecker:2021}. We respond to this literature by pointing out the
cultural layers which may interfere with the neutrality and
machine-readability of visualization across disciplines.

\subsection{Studies of digital media objects}

There has been growing interest in analyzing code
and its visual results as media objects: artifacts which trace the 
development of a media culture over time, like wax cylinders or old 
television sets. Jacob Gaboury's \emph{Image Objects: An Archaeology of
Computer Graphics}~\cite{gaboury2021image} takes a media archaeology approach to
five such objects in the history of computer graphics: the hidden surface
problem, the frame buffer, the virtual teapot, object orientation, and
the GPU. In \emph{10 PRINT CHR}~\cite{montfort201410}, Montfort et
al.\ study a line of BASIC code used to generate a random maze and use
it as an entry-point to explore the cultural history of the Commodore
64 as well as topics including mazes, grids and randomness. These works live within the space charted by Lev Manovich's
now-canonical \emph{The Language of New Media}~\cite{manovich2002language}, 
which frames ``new media'' as visual artifacts produced by textual code, and argues that new media demands new theoretical perspectives.

%There is also a rich history of ``distant reading'' (i.e. looking at a
%corpus of texts in aggregate, usually with computational methods) in
%the digital humanities. For example, Moretti studies the titles of novels published
%in Victorian England~\cite{moretti2009style}. Goldstone and Underwood
%use topic modeling to investigate the changes over time in a large
%corpus of articles from literary studies~\cite{goldstone2014quiet}. Arnold and Tilton describe a theoretical
%framework for extending distant reading to ``distant viewing'' of
%visual culture which they apply to historical photographs and
%television \cite{arnold2019distant}. While we rely primarily on
%qualitative analysis, the concept of distant reading and viewing is central to our
%approach.

\subsection{The research paper as a site for interaction design research}

Closer to CSCW, there are several studies of the research paper as a
site for social computing research and designs for digital literature reviewing: Chang et al.\ introduce CiteSee, a tool for visually augmenting citations in PDF documents based on the reader's prior research activity \cite{chang2023citesee}. Qian et al.\  describe commercial tools for PDF management as ``iTunes for papers,'' as they treat the research paper as a fundamental unit of scholarship, and propose a literature review tool to break down papers further into collections of specific grounded claims \cite{qian2019beyond}. 

This research community is also interested in designing for the writing process: Head et al.\ study ways of
augmenting digital documents with definitions of terms and symbols to
improve readability~\cite{head2021augmenting}, and how authors improve
readability by augmenting the visual design of their
equations~\cite{head2022math}. Manzoor et al.\ develop a LaTeX editor
extension to improve the accessibility of LaTeX for authors with visual
impairments \cite{manzoor2018assistive}, and Hara
et al.\ develop a system for generating Braille documents from
mathematical expressions written in
LaTeX~\cite{hara2000mathbraille}. Gobert and Beaudouin-Lafon conduct a
study of LaTeX users and design an extension for VSCode that uses
transitional representations of document objects like tables to
improve the editing experience~\cite{gobert2022latex}.  Haber et
al.\ study how groups of coworkers interact differently when using
physical vs. virtual documents~\cite{haber2014paper}.  While we do not design or develop any technical
tools, our work provides additional evidence for the importance of further design studies of research paper reading and writing.

\subsection{The culture of computer vision}

Motivated by the dangers of algorithmic discrimination and safety
concerns in systems relying on computer vision algorithms, several
recent papers have studied the culture of the computer vision research
community and its understanding of data and truth, often using
research papers as texts for analysis. Su and Crandall study the
emotional state of the computer vision community, finding that the
deep learning revolution and subsequent growth have had a profound
effect on its culture,  leading   both to excitement  as well as  isolation and
malaise~\cite{su2021affective}. Denton et al.\ use a discourse
analysis approach to study the history of the ImageNet dataset through
the research papers and presentation slides of Fei-Fei
Li~\cite{denton2021genealogy}. Scheuerman, Denton, and Hanna study a
corpus of 500 papers describing computer vision datasets and analyze
the values implicit in their writing: 
efficiency, universality, impartiality, and model
development over dataset development~\cite{scheuerman2021datasets}. While our work also analyzes computer vision though its research papers, our interest is in studying these papers as primarily visual media.

A wide variety of authors have written critically about the culture of
data collection and dataset use in machine learning. For example,
Sculley et al.\ and Ethayarajh and Jurafsky critique the concept of
benchmarks and leaderboards in machine learning, arguing in different
ways that steadily increasing scores do not always correspond to
progress~\cite{sculley2018winner,ethayarajh2020utility}. For a more
comprehensive survey, please see Paullada et
al.\ \cite{paullada2021data}. Our work here, which does not investigate
data collection or dataset usage directly, nevertheless echoes these
themes.

\section{Methods}

Our inquiry began through visual analysis of historical computer
vision research papers. We were surprised by visual design changes across 
these papers, particularly the increasing prevalence of
highly complex figures, and decided to investigate further by 
conducting  semi-structured interviews with researchers who had been
active in computer vision, graphics, and visualization for several
decades, as well as computational analysis of a corpus of
research papers published between 2013 and 2021 in the IEEE/CVF Conference
on Computer Vision and Pattern Recognition---the highest-ranked publication venue in computer science and second highest in all of science, according to Google Scholar as of August 2024 \cite{googlescholar}. We study this conference, rather than similar conferences in machine learning or natural language processing, because of the importance of visual presentation in computer vision. Along the way, our visual analysis, interviews, and computational findings coalesced into a media archaeology approach.

\subsection{Media Archaeology}

Inspired by recent work on the history of computer graphics
\cite{gaboury2021image}, we examine historical research papers through
the lens of media archaeology. ``Archaeology'' here refers to Foucault's concept of archaeology as the ``uncovering of the archive'' \cite[p. 131]{foucault1969archeologie}, rather than physical excavation. Media archaeology, as defined by Huhtamo and Parikka, ``rummages textual, visual and auditory archives as well as collections of artifacts, emphasizing both the discursive and material manifestations of culture'' \cite[p. 3]{huhtamo2011media}. In other words, we study digital media objects as components of a media culture: a system of practices and meanings which structure our interpretation~\cite{rose2016visual}. We also consider their digital materiality as physical patterns of bits on hard drives, produced through the labor of researchers. By drawing attention to material culture, media archaeology is in dialogue with theories of historical materialism, which consider economic factors---production and consumption of goods and services---as the driving force of history, rather than revolutionary ideas or progress narratives \cite[p. 5]{gaboury2021image}.

To illustrate these concepts, consider Figure 2(d), a teaser image from a paper in the
proceedings of the IEEE Winter Conference on Applications of Computer
Vision (WACV) \cite{kong2022hole}. This figure is split into 5 similar
subfigures, with different combinations of black lines and teal
dots. With the requisite disciplinary knowledge, a viewer can
interpret the meaning of the figure: the five subfigures represent an
occluded image and four attempts to reconstruct the wireframe of the
original using different automated methods. Using our understanding of
the task, we clearly see that the fifth subfigure labeled ``ours'' is
the best, because its dots and lines align with the geometry of the
depicted room.
These inferential steps 
require a degree of initiation in computer vision, an understanding of
both the goal of a research paper and the system of meaning that these
papers use. The goal of media archaeology is to dissect these
inferences and identify how their layers of meaning developed over time.

In addition to our computational analysis and detailed study of the CVPR proceedings, we ``rummaged'' research papers from the IEEE, Computer Vision
Foundation, and ACM digital archives, including the IEEE International
Conference on Computer Vision (ICCV), European Conference on Computer
Vision (ECCV), IEEE Transactions on Pattern Analysis and Machine
Intelligence (PAMI), and the ACM Special Interest Group on Graphics and
Interactive Techniques conference (SIGGRAPH). We examined ICCV,
ECCV, and PAMI in addition to CVPR because they are the four most influential computer
vision venues according to Google
Scholar,\footnote{\url{https://web.archive.org/web/20220709182958/https://scholar.google.com/citations?view_op=top_venues&hl=en&vq=eng_computervisionpatternrecognition}}
and SIGGRAPH due to its historical significance and
prevalence in our interviews. We turned to these venues to build context for the results we were seeing in CVPR. While we did not systematically code the
enormous number of papers published in these venues, we collected
dozens of screenshots of interesting figures and tables and discussed
their visual style in weekly meetings, integrating interview and
computational results which eventually coalesced into historical
narratives.

\subsection{Interviews}

To augment our visual analysis, we conducted
interviews with veteran computer vision researchers. As there are a
relatively small number of eligible participants, we recruited individuals via email and in-person at the CVPR 2022 conference. Participants are listed by years of industry and academic experience in Table \ref{tab:participants}. This sample is highly non-representative, showing both survival bias
based on who remains in the field over a long period of time and
selection bias based on who agreed to an interview, so it does not serve as the sole source of our claims. Instead, it provides important social context to our visual analysis.

We conducted interviews in person and over the Zoom videoconferencing platform between March and August 2022. To avoid participants historicizing or theorizing themselves, we asked each participant to discuss a specific research paper from their early career, and asked them to explain the different elements and tell us stories about the writing process for that paper. Additionally, we asked each participant how they read research papers at that time of that paper, and how those reading and writing processes differ from those of the present. This method of interviewing mirrors that of previous studies examining media artifacts such as websites~\cite{ChenCrandallEtAl:2017,goree2021investigating}. 

After transcribing the interviews, we engaged in a process of
iterative memoing, integrated into our weekly meetings where we compared examples from historical research papers and
interview excerpts which shared common themes. We focused our analysis on
identifying patterns from the context that participants provided while
telling stories about their writing, as well how  participants
thought about their papers in relation to changing disciplinary
practices, rather than simply analyzing the papers participants discussed or discussing our participants' analysis of their own work.

\begin{table}[]
    \centering
    \begin{tabular}{cccc}
        \toprule
        Participant & Discipline & Years Industry & Years Academia \\
        \midrule
        P1 & CV,G & 9 & 33\\
        P2 & G,V & 3 & 17\\
        P3 & CV & 0 & \textbf{42}\\
        P4 & CV & \textbf{21} & 3\\
        P5 & CV & 15 & 11\\
        P6 & CV & 0 & 10\\
        P7 & CV & 0 & 36\\
        P8 & CV & 0 & 16\\
        P9 & CV & 0 & 19\\
        P10 & CV,V & 0 & 22\\
        P11 & V & 11 & 0\\
        P12 & CV,G & 0 & 23\\
\bottomrule
    \end{tabular}
    \Description{10 participants come from computer vision, 3 from graphics and 2 from visualization. 11 have academic experience, ranging from 10 to 42 years, 5 have industry experience ranging from 3 to 21 years. Facetiously, the highest number in each column is bolded.}
    \caption{List of participants. Discipline is some combination of computer vision (CV), graphics (G), and visualization (V). Years in industry and academia are defined as years since PhD spent employed by a university or an industry research lab. Joint appointments are counted as academia for simplicity. 10 identify as men and 2 identify as women, and 7 identify as white and 5 identify as Asian.}
    \label{tab:participants}

\end{table}

\subsection{Computational Analysis}

As both qualitative approaches rely on the analysis of specific
examples, we used supplementary quantitative analysis to verify that
phenomena we observed in the interviews and media archaeology were actually as pervasive as participants seemed to think. Specifically, we were interested in whether teaser images, figures and tables were becoming more prevalent in CVPR papers over time, and whether more titles were following a particular format containing an acronym followed by a colon. To answer these questions, we collected PDFs of papers from the Computer Vision Foundation's Open Access Repository 
(\url{thecvf.com}) published in CVPR 2013 to 2021 and article metadata from IEEE Xplore
(\url{ieeexplore.ieee.org}) for CVPR 1992 to 2020. 
We manually inspected PDFs from \url{thecvf.com} to count teaser images, then used the Linux \texttt{pdftotext} tool combined with regular expressions to count figures and tables. From the Xplore metadata, we used regular expressions to parse paper
titles, treating a word with more than two capital letters as an
acronym, and an acronym as unique if it only appears once in a given
year of data.

\section{Findings}

\begin{figure*}
    \centering
    \begin{tabular}{c c c}
        \includegraphics[trim=0.1in 4.5in 0.1in 0.8in,clip,width=0.25\textwidth]{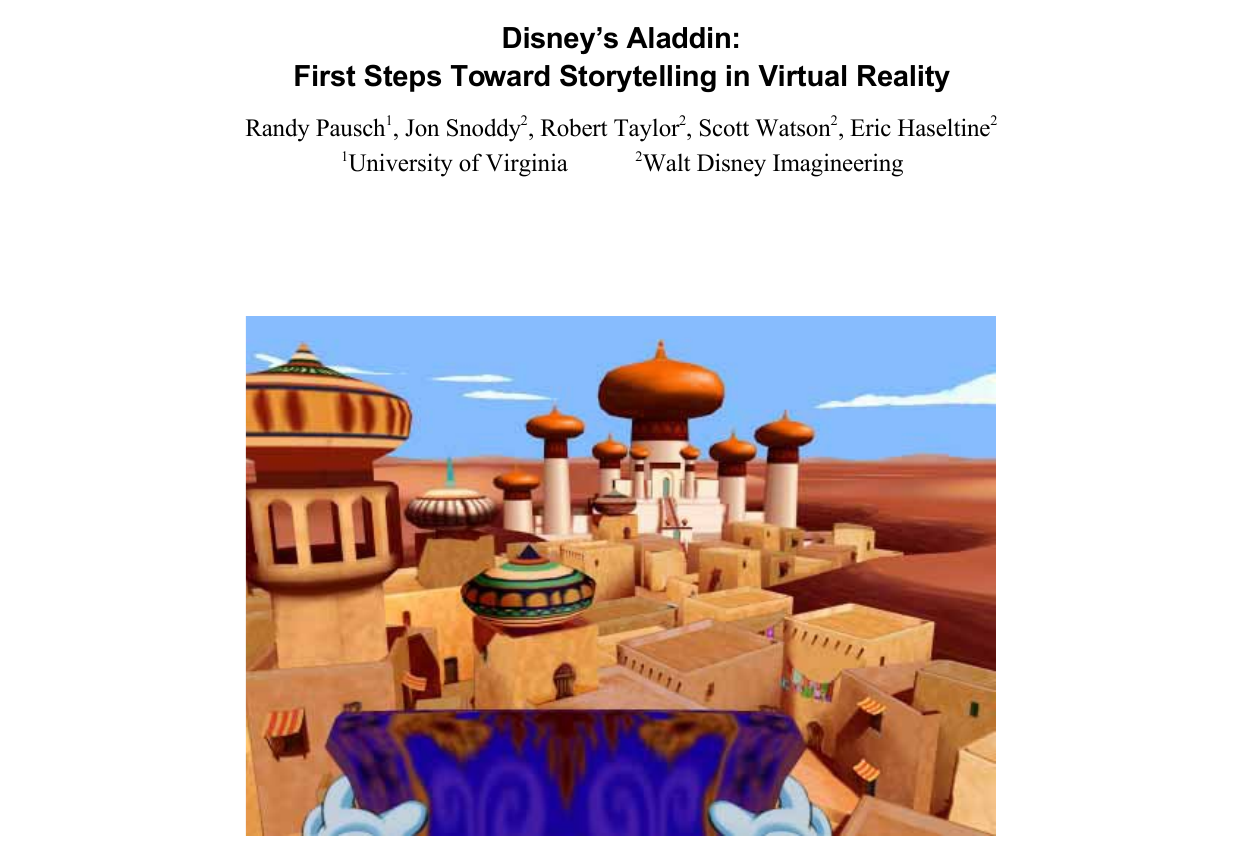} & 
        \includegraphics[trim=0 4in 0 0,clip,width=0.25\textwidth]{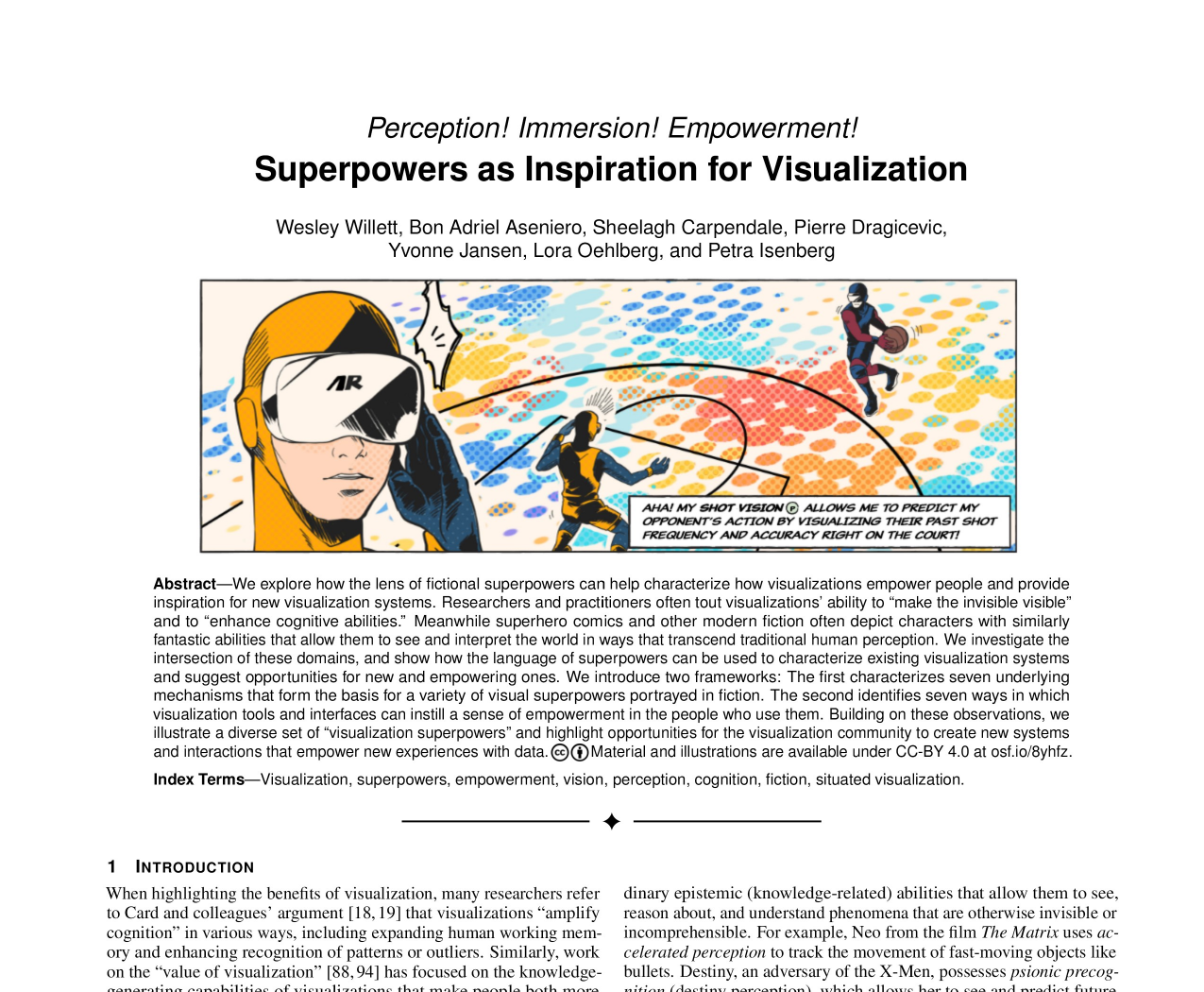} & 
        \includegraphics[trim=0 4in 0 0,clip,width=0.25\textwidth]{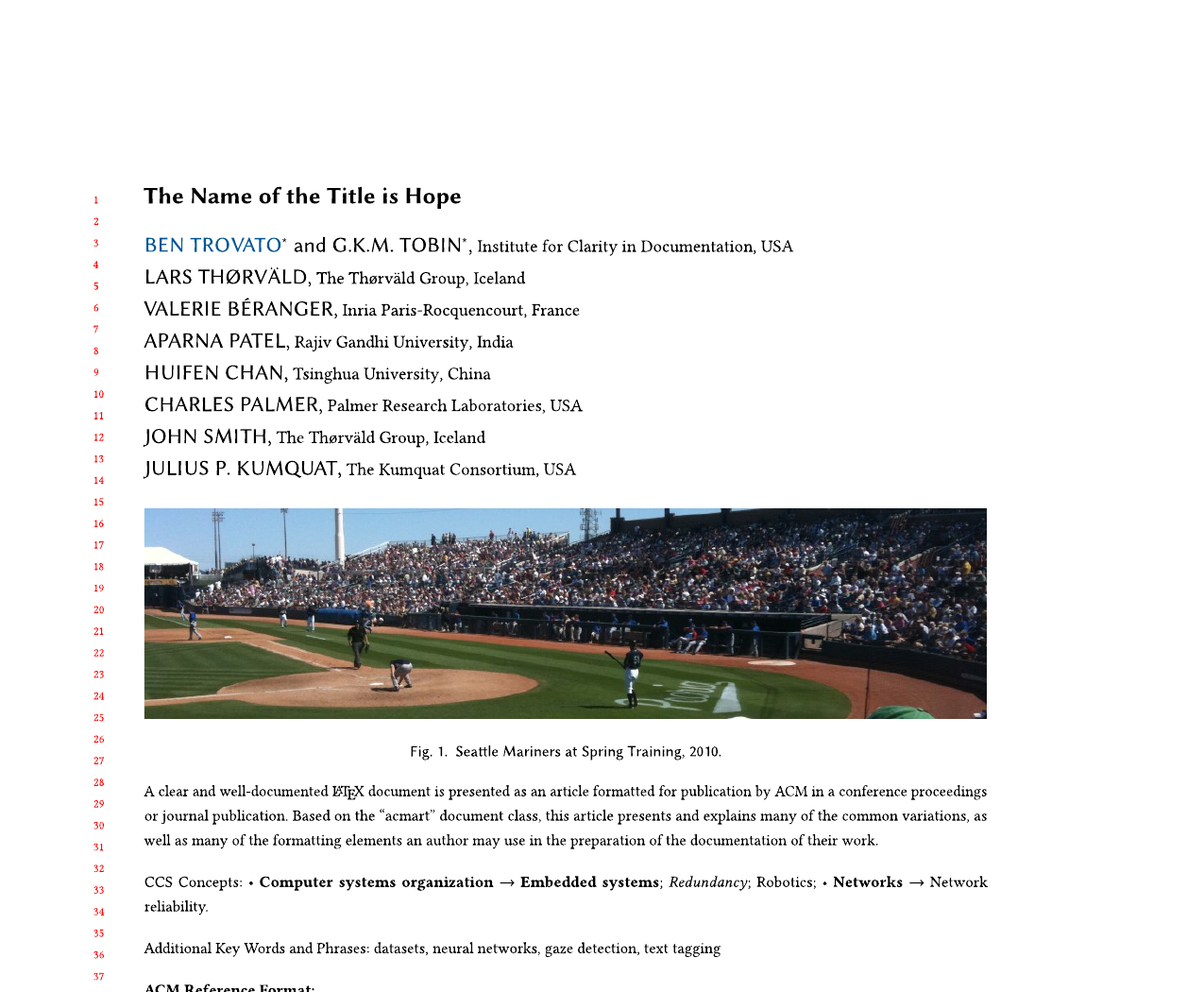} \\
        (a) SIGGRAPH 1996 \cite{pausch1996disney}& 
        (b) TVCG 2021 \cite{willett2021superpowers}& 
        (c) The ACM conference template \\ \\
        \includegraphics[trim=1in 6in 1in 1in,clip,width=0.25\textwidth]{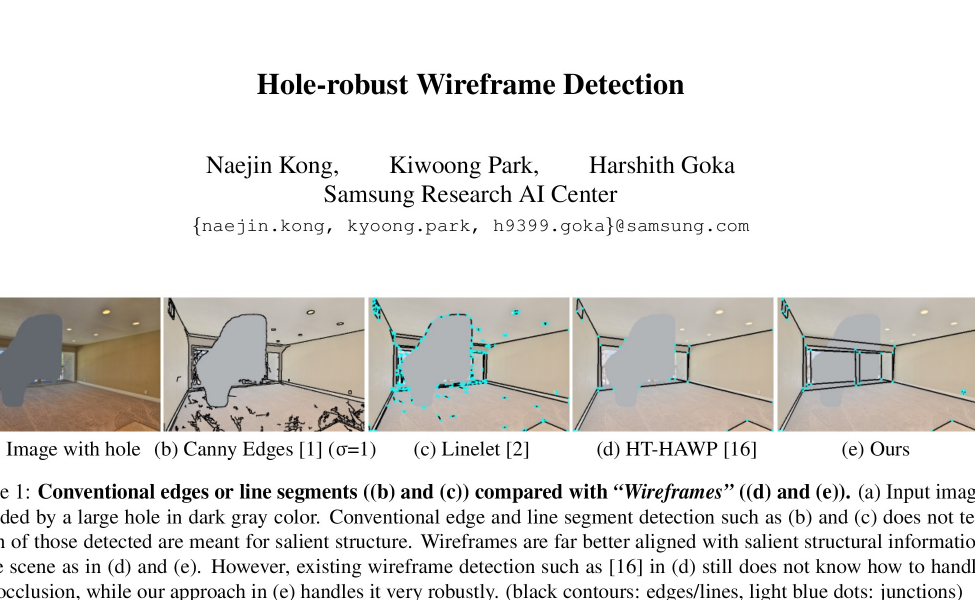} & 
        \includegraphics[trim=0.5in 5.3in 0.5in 1in,clip,width=0.25\textwidth]{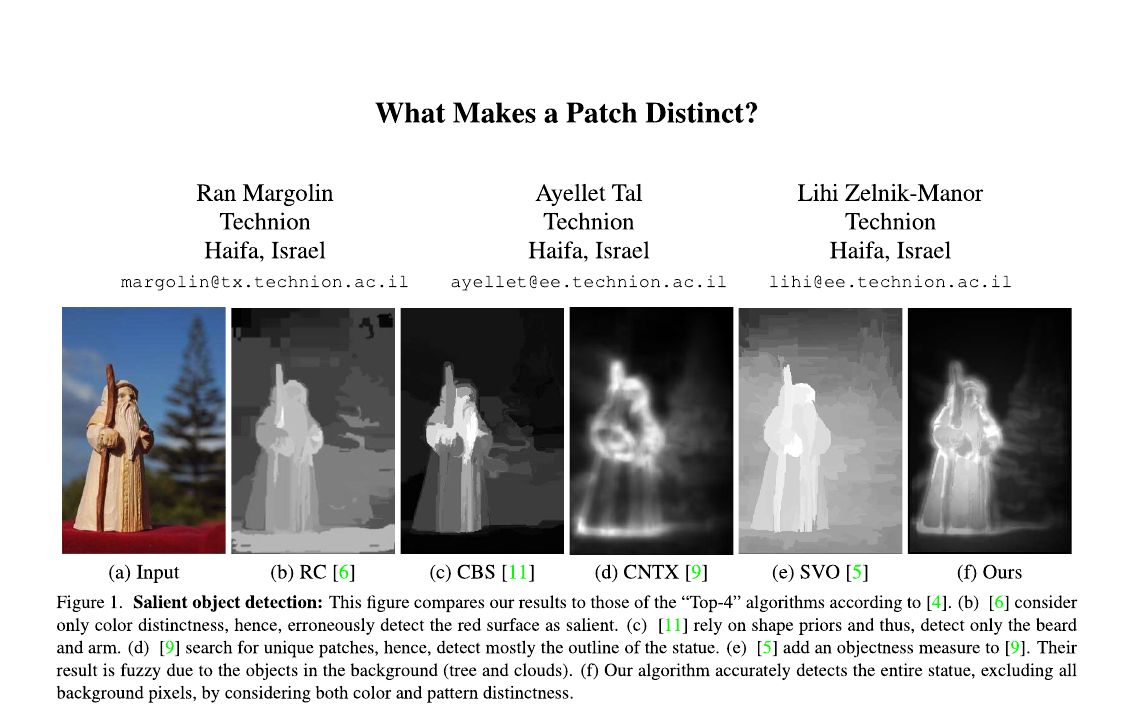}&
        \includegraphics[width=0.25\textwidth]{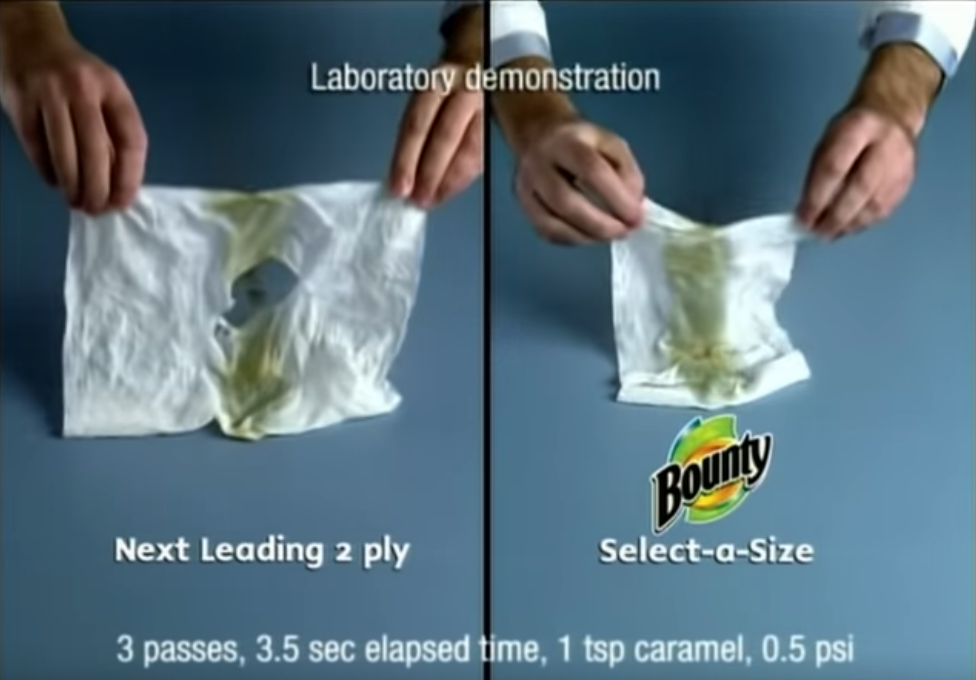}\\
        (d) WACV 2022 \cite{kong2022hole}& 
        (e) CVPR 2013&
        (f) A paper towel ad from the early 2000s \cite{bountyad}.
    \end{tabular}
    
    \caption{Five teaser images from papers in different venues, and a
      still image from a television advertisement for paper
      towels. Figures look best zoomed in.}
    \label{fig:teasers}
    \Description{Six subfigures. (a) shows Randy Pausch's 1996 teaser image, a large color figure of an Aladdin-themed virtual reality ride. (b) shows a 2021 visualization paper from TVCG which uses superhero imagery to illustrate its design approach. (c) shows the ACM conference template, with its stock baseball teaser image. (d) shows a paper from WACV 2022 with a teaser showing an image of a room containing a missing image region and four attempts to locate the edges of the room. The last method labeled ``ours'' clearly has the best results. (e) shows an input image of a small figurine, possibly Moses, and five competing methods to predict saliency maps over the image. Again, the last method, marked ``ours'' does best. (f) a still from a paper towel TV advertisement showing ``Bounty select-a-size'' effectively cleaning up a spill while ``Next leading 2 ply'' fails. The image also reads ``Laboratory demonstration. 3 passes, 3.5 sec elapsed time, 1 stp caramel, 0.5 psi.''}
\end{figure*}

Through our analysis, we found that several elements of the contemporary computer vision research paper serve as material traces of the disciplinary change which took place during the 2010s. We find that papers are written to promote a specific contribution which is increasingly commodified. By ``contribution,'' we mean the model, algorithm, method, dataset, or other system that the paper offers to readers. By commodified, we mean reduced to its exchange value in terms of the attention it generates. As Simon theorized \cite{Simon1971}, and authors including Goldhaber \cite{goldhaber1997attention} and
Davenport and Beck \cite{davenport2001attention} have applied to the web, when ``capital, labor,
information and knowledge are all in plentiful supply'' the limiting
factor is the attention of consumers,
which begins to be treated like a commodity \cite{davenport2001attention}.

However, this concept of the attention economy presupposes that attention is valuable, which is unhelpful for understanding why academics seek the attention of their peers instead of the larger research community or general public. To better understand why academic attention is uniquely valuable, we employ Bueno's explanation, that online attention, in addition to a commodity, is a form of labor \cite{buenoattention}. This claim relies on Romano Alquati's concept of ``valorization information'' via Pasquinelli \cite{pasquinelli2015italian}. Alquati, in his study of an Olivetti computer factory, found that workers contribute to mostly-automated production processes through their micro-decisions, which served as a feedback-control signal for the production process. Bueno argues that attention can provide a similar signal. Specifically, by monitoring attention, companies can forecast consumer behavior and more efficiently follow market trends. By doing so, these companies extract surplus value from online behavior data \cite[Ch.\ 2]{buenoattention}. We believe a similar process may be at play in academic publishing, where authors can leverage highly specialized early attention to increase the prestige value of their research. This process can take the form of feedback, where authors improve their work based on an early online response, or simply where early viewers find a paper interesting and share it, judging it as valuable, putting it in front of more eyes and increasing the chance that it will reach the right audience. Eventually, that attention translates into citations, jobs, students, grants, and other benefits of academic prestige.

We study three aspects of contemporary research papers:  the teaser image, the results table, and the high resolution figure. We have grouped our results into thematic sections surrounding these three elements in relation to the concept of a paper's contribution.

\subsection{Advertising the Contribution}

\subsubsection{Teaser images}

\begin{figure*}
    \centering
    \includegraphics[width=0.4\textwidth]{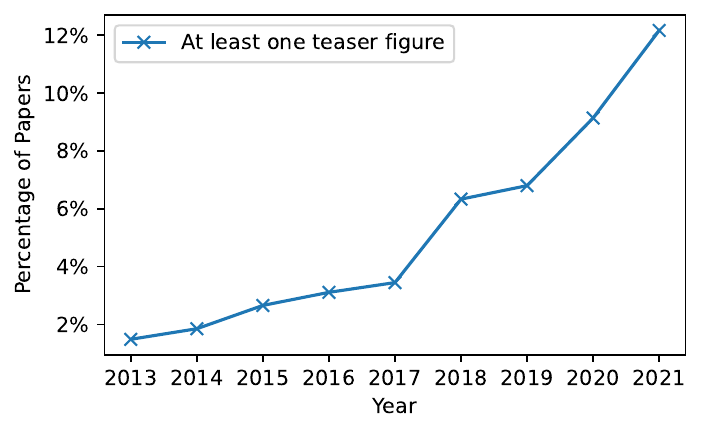}
    \includegraphics[width=0.4\textwidth]{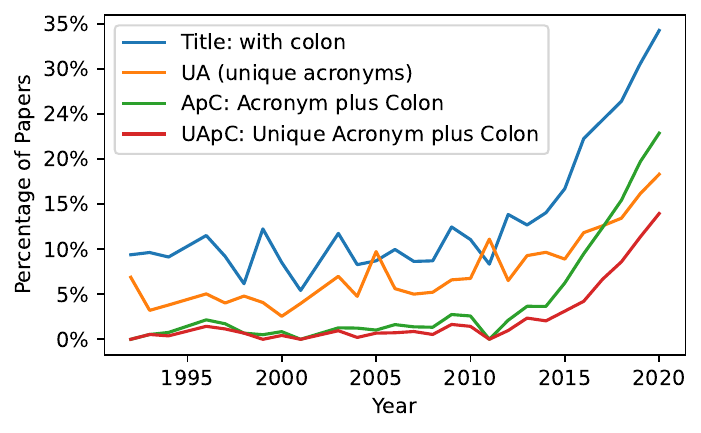}
    \caption{Left: fraction of CVPR papers with a teaser image. Right: fraction of CVPR paper titles  with colons, unique acronyms, acronyms  followed by a colon, and unique acronyms followed by colons. Differing timescales are due to differing availability of full PDFs vs. title data. \vspace{-10pt}}
    \Description{Two subfigures. Left shows a graph with ``fraction of papers with teaser figures'' on the Y axis and ``Year'' on the X axis. The line increases from 2\% in 2013 to 12 \% in 2021. Right shows a graph with four lines, representing four quantities which increase between 1992 and 2020: ``Title: with colon'' increases from 10\% to 35\%, ``UA (unique acronym)'' increases from 7\% to 17\%, ``AFbC: Acronym Followed by Colon'' increases from nearly 0\% to 22\% and ``UAFbC: Unique Acronym Followed by Colon'' increases from nearly 0\% to 15\%.}
    \label{fig:figure_size}
\end{figure*}

The ``teaser image'' is a large  first-page figure which summarizes the paper. Several examples of teaser images are shown in Figures \ref{fig:teaser} and \ref{fig:teasers}. These figures are functionally similar to the trend of ``visual abstracts'' \cite{ramos2020visual} and ``table of contents images'' \cite{buriak2011summarize} in the biomedical sciences; however unlike these forms which are graphical summaries of the text, teaser images are part of the main paper and are usually a visualized system output that the authors want to show. Teaser images have been steadily gaining popularity at CVPR, as shown in Figure \ref{fig:figure_size} (left). For this analysis, we define the teaser image as a first-page figure which covers the entire width of the page, not just a column.

P12 ascribes the teaser image to the famous graphics researcher Randy Pausch: \textit{``Randy Pausch...takes credit for putting the teaser
  image in SIGGRAPH papers...he claimed that he did a paper where half
  the first page was a teaser image and after that that became the
  norm where people started always putting these images.''}  The paper
in question is pictured in
Figure~\ref{fig:teasers} (a). While Pausch did not invent the large first-page
figure (Tim Berners-Lee famously used one in his 1989 proposal for the
world wide web \cite{berners1989information}), he did publish the first
paper in SIGGRAPH with this layout, and it quickly became popular. P2
and P10 point to the quick adoption of teaser images, which even
became institutionalized in the SIGGRAPH template and later the template for all ACM conferences (Figure \ref{fig:teasers} (c)).

P12 defines the teaser as a visualization of either a result or a
system and explains that it has spread to many other 
conferences because of the way it attracts attention: \textit{``It's a
  trailer. It's to get people in...I think it's a very compelling way
  to convey what the paper does.''} For her, the teaser is a highly
  effective innovation which improves research papers. P2 echoes 
  that sentiment:
\textit{``They just made the papers look good! I mean, it's much more
  memorable and there are some papers still today that I don't
  remember the title, but you see the picture and you're like, `Oh
  yeah, that's the Randy Pausch paper on the VR for whatever, right?' 
  So yeah there's a few of those that are just like, really iconic
  first pagers.''} P2 is referencing the same Pausch paper as P12 \cite{pausch1996disney}, highlighting its memorability.

The teaser image is a trailer, a \textit{hook}; it advertises the
paper to the potential reader. The authors want to promote their paper and 
showcase the best results they can because the sheer number  and easy 
access to papers has made it harder to stand out. The visual organization 
of these figures echoes
that theme. Notice the commonalities between the two teaser images in
Figure \ref{fig:teasers} (d) and (e). Both show the output of several
algorithms attempting to solve the same problem, but they depict the
results in such a way that their method is clearly best. This visual
effect mirrors that of advertisements, which depict two competing
products in action (Figure~\ref{fig:teasers} (f)). While the
experiments in research papers are far more sophisticated than those in advertisements, they share a common visual design. Along these
lines, we find a recurring theme of teaser images which rely on the
iconography of the 3D-reconstructed, brightly colored human body as a
particularly compelling visual. Several examples are shown in Figure
\ref{fig:teaser}. Human faces and bodies have been found to attract
consumer attention to advertisements
\cite{wilkinson2011preliminary,ohme2011biometric}; a similar principle
may be motivating computer vision authors.

We observe a similar advertising quality in titles. Figure
\ref{fig:figure_size} (right) shows the rise in popularity of a
particular title construction where an acronym, which is usually the
name of a model, is followed by a colon: ``HOPE-Net: A Graph-Based
Model for Hand-Object Pose Estimation'' \cite{doosti2020hope} or
``DeMoN: Depth and Motion Network for Learning Monocular Stereo''
\cite{ummenhofer2017demon}. These names both signal that the paper
proposes a new model and brands the paper with a short, memorable name
which is often a cultural reference or clever pun. An especially
cheeky example is Joseph Redmon and Ali Farhadi's 2017 paper
``YOLO9000: Better Faster Stronger''~\cite{redmon2017yolo9000}, which
improves upon an earlier model called YOLO (short for
You Only Look Once)~\cite{redmon2016you}, and references a variety of
memes from the mid-2010s.\footnote{YOLO itself was a meme, short for
  ``you only live once,'' the number 9000 references a famous quote
  from the television show Dragon Ball Z and the phrase ``better,
  faster, stronger'' references a Daft Punk song. While unusual in
  computer vision, this kind of nerdy referential humor has been
  observed in other areas, for example, web design~\cite{goree2022really}.}

\subsubsection{Beyond the paper: arXiv, social media and videos}

Why would many authors begin to engage in these teaser figure and title practices over the course of the 2010s? One explanation is  a change in medium through which research reaches its audience, from conference and journal proceedings onto social media and preprint websites. P4 describes how his students have strategies
to get their papers noticed on the preprint site arXiv: \textit{``if they take your submission on Thursday and it goes up on the arXiv most recent publications list, it's up for all of Friday, Saturday, and Sunday...so it gets more days of exposure...It improves your odds that somebody will notice your thing.''} P8 describes the pressure on her students to promote their work:
\textit{``social media, like promoting one's research has become such
  a big thing and I think students...are realizing, `Oh it's not enough
  for me to, y'know, come up with a paper, post it on arXiv, get it
  accepted. I also need to tweet about it.' It's frankly quite
  exhausting...[before] the only way you promote your paper is it
  shows up at the conference and hopefully, y'know, some famous
  computer vision researchers will come up and look.''}  She talks
about how her advisor would bring his friends over to her poster, and
that was all the advertising she needed to get noticed, but that
strategy no longer works as well in the modern crowded field of computer
vision.

Again this pattern resembles one from computer graphics. P12 describes 
the importance of videos in SIGGRAPH: \textit{``in
graphics the conventional wisdom has been it's useful to have, not
only do the paper...but to also show a video that really highlights
and explains the work. And if you do a good job of explaining it,
people find it compelling.''} These videos were considered part of
the paper submission and peer reviewed alongside the text, which led
authors to invest heavily in the quality of their videos.

That attention translates into citations, as well as other benefits. P11, who works in an industry lab that hires recent PhDs, explains why his group publishes at all: \textit{``exposure is actually really important. If you want to attract really top level talent, then having zero published papers is really going to work against you, right? Particularly if you're looking at people who are in positions that are for doing some sort of active innovation.''} For this group, the publication is primarily a way to gather attention, they derive value from the attention paid to their work through hiring. Second, attention helps researchers to identify the value of their own work. P7 talks at length about his desire for impact. He switched fields from computational complexity to computer vision because there were more active researchers, and he could not tell if his work was having an impact in a small field. P9 tells the story of how attention from a senior scholar improved his relationship with his advisor: \textit{``that email from [scholar], that’s the beginning of my great relationship with my advisor was that one little email, because within five minutes [advisor] was up in the office, `Oh, you got an email from [scholar]? Very good. Very good. How about a coffee?'\thinspace''} Notice that the career benefits resulted not from the attention itself, but from the record of that attention in the form of an email from a senior scholar.

While the relationship between attention and prestige is not new, the unprecedented growth of computer vision over the past decade places pressure on the attention of senior researchers. There have been ongoing discussions in the computer vision community on how to react to this growth. Some of the proposed ideas are recorded in the passed motions of the annual meeting of the IEEE pattern analysis and machine intelligence technical committee (PAMI-TC), the governing body of the computer vision research community. In 2021--2022, motions changed the rules regarding double submission to reduce burden on reviewers, introduced punitive measures for missing or poor quality peer reviews from authors, and banned discussion of submitted papers on social media to avoid influencing peer reviewers (the last of which was repealed in 2023) \cite{pamitc,pamitc2021}. The rationale statement for the social media ban specifically highlights attention: ``Groups with large followings and the resources to mount visible social media promotions received significant attention for work that is under review. Reviewers are exposed to this work and the attention it receives can bias their judgment---if so many people on social media are excited, mustn't it be good?'' \cite{pamitc2021}. In addition to highlighting the valorizing role of attention, this statement showcases the view of peer review as an unbiased process for evaluating the quality of research. 

We see two additional key themes starting to emerge. First, there are
parallels between computer vision in the 2010s and computer graphics in the 1980s and 1990s,
 which many of our participants pointed out (P1, P2, P6, P9,
P10, P12). Both disciplines rapidly grew  due to
industry investment, from the tech industry in vision and the
entertainment industry in graphics, which created an attention
economy, forcing papers to go above and beyond to attract attention. Second,
research work is being commodified---treated as
interchangeable, given some measure of its value. In computer vision,
that means the particular ideas an author proposes are less important
than their ability to grab attention and advance the author's career.

\subsection{Measuring the Contribution}

\begin{figure*}
    \centering
    \includegraphics[height=4cm]{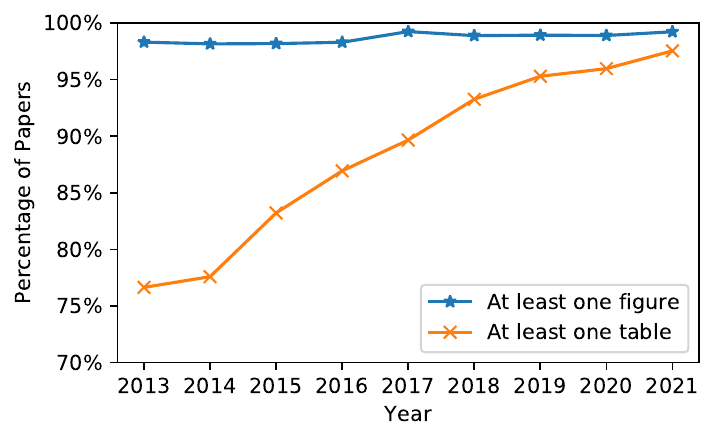}
    \includegraphics[height=4cm]{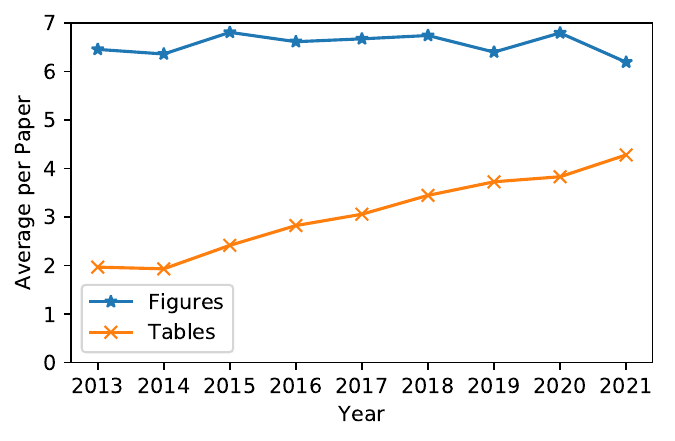}
    \caption{Left: The fraction of CVPR papers with figures and tables over time. Right: The average number of figures and tables per CVPR paper over time.}
    \label{fig:table_counts}
    \Description{Two subfigures, both line plots. Left shows the percentage of papers with a figure remaining constant around 99\% while the percentage of papers with a table increases from 77\% in 2013 to 97\% in 2021. Right shows the average number of figures and tables per paper. Figures are constant between 6 and 7 while tables increase from 2 in 2013 to 4 in 2021.} 
\end{figure*}

Today, the ``table of results'' in a computer vision paper fulfills a central function in both the
written argument  and the peer review process:
evaluation. When proposing a new method for solving a vision problem,
the authors must demonstrate that it works at least as well, if not
better than, ``SOTA'' (state-of-the-art) 
existing methods. These
tables often contain the values of standard evaluation metrics computed
on a benchmark dataset, like top-1 or top-5 accuracy on the ImageNet
test set~\cite{deng2009imagenet} for image classification or mean
average precision on the MS-COCO test set~\cite{lin2014microsoft} for
object detection.

\begin{figure*}
    \centering
    \begin{tabular}{c c c}
    \includegraphics[width=0.25\textwidth]{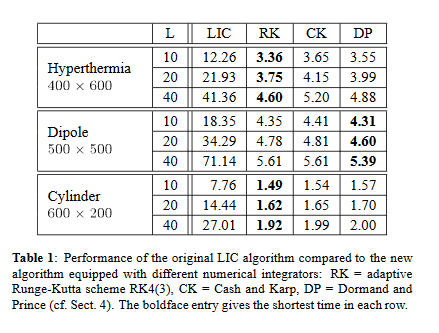} &
    \includegraphics[width=0.25\textwidth]{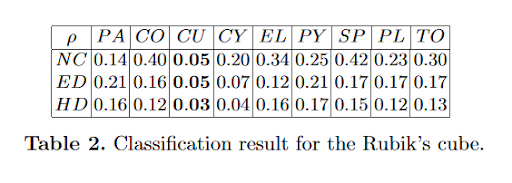} & \includegraphics[width=0.25\textwidth]{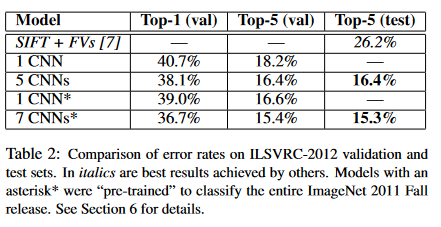}\\
    (a) SIGGRAPH 1995 \cite{stalling1995fast} & (b) 1999 edited volume \cite{chella1999cooperating} & (c) NeurIPS 2012 \cite{NIPS2012_c399862d} \\ \\
%    \end{tabular}
%    \begin{tabular}{c c c}
    \includegraphics[width=0.25\textwidth]{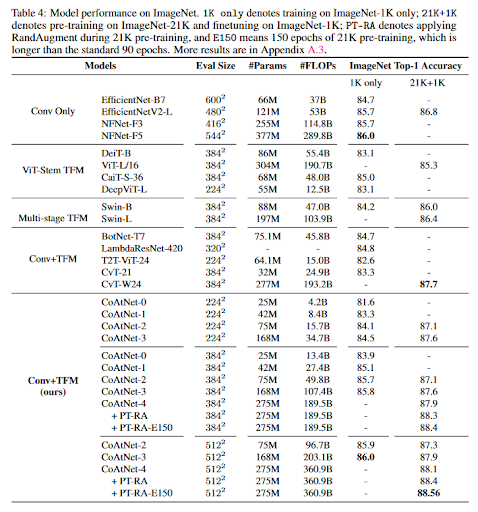} &
    \includegraphics[width=0.25\textwidth]{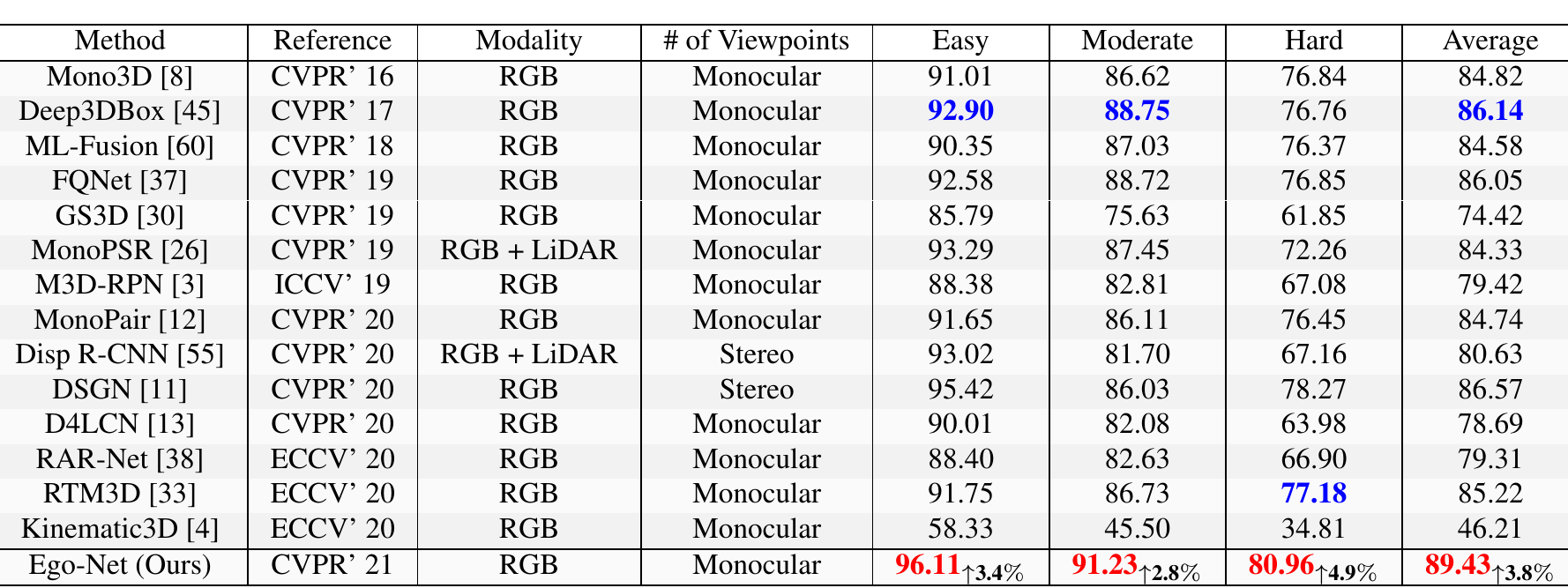} &
    \includegraphics[width=0.25\textwidth]{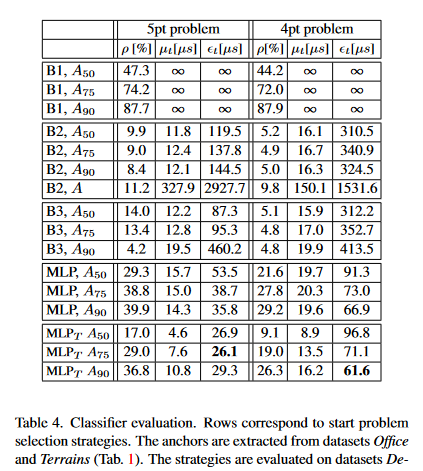}\\
    (d) NeurIPS 2021 \cite{dai2021coatnet} & (e) CVPR 2021 \cite{li2021exploring} & (f) CVPR 2022 \cite{hruby2022learning}
    \end{tabular}
    \caption{Six results tables with numbers in bold. (a) is the earliest example of this style we found, (b) is the earliest example from computer vision. (c) is from the highly influential 2012 AlexNet ImageNet classification paper \cite{NIPS2012_c399862d}, (d) is a 2021 state of the art result on ImageNet \cite{dai2021coatnet}, (e) is a more trendy table from 2021, making use of grayscale background, colored numbers and subscript arrows showing improvement \cite{li2021exploring} and (f) is a table from a 2022 CVPR paper \cite{hruby2022learning} from the geometric side of computer vision, which is historically more mathematical and usually has fewer such tables.}
    \label{fig:table_examples}
    \Description{Six subfigures. Each shows a table full of numbers, where some of the numbers are in bold. (a) shows a table split into sections with double lines where one number in each row is bolded. (b) shows a simpler table where all of the numbers in one column are bolded. (c) shows another simple table where one row has an italicized label and a citation, while two of the other rows have numbers in bold. (d) shows a highly complex table with dozens of rows split into several sections with horizontal lines and several numbers in bold. (e) shows a dense table with a grayscale background, alternating in tone to differentiate the lines, and numbers in both blue and red. (f) shows a table with many rows split into sections with double horizontal lines, two numbers in bold and several cells with infinity symbols.}
\end{figure*}

While it is tempting to treat these tables as textual content, their effect on the paper, we argue, is primarily visual. A key feature of these tables is that they put the best result, which
is almost always from the author's proposed method, in bold. This
design feature is essential for readability, as a large table full of
numbers is very difficult to interpret. These tables will sometimes
also use arrows to indicate whether a column displays a metric where
higher or lower numbers indicate better performance. More recently, as
these tables have become more complex, authors have developed other
readability innovations, like using colored numbers and subscripted or
parenthetical percent improvements (Figure
\ref{fig:table_examples} (e)). These innovations serve to further visualize the quantified value of the paper's contribution.

But all of this was not the case a few decades ago. The vast majority
of computer vision papers from the 1980s and 1990s rely on
mathematical arguments based on, for example, pinhole camera geometry and do not
contain any quantitative results. Empirical evaluation, if included at
all, was primarily qualitative, in the form of figures showing sample results.
As P8 explains, she had a combination of
quantitative and qualitative evaluation in her paper from 2003, which
was unusual: \textit{``quantitative evaluation, you know, back in 2003
  was still kind of in its infancy...I'm not sure that this [2003] paper has
  basically any comparison to competing methods which probably would
  be required today.''} P3 explains that he was primarily concerned in
1990 with showing test examples to demonstrate his algorithm's
effective handling of edge cases. P9 explains that in 1999, showing
example output of his system was sufficient: \textit{``instead of
  [Amazon] Mechanical Turk you just have the reviewers just eyeball the
  images.''} In the satirical 2010 ``Paper Gestalt''~\cite{von2010paper} paper,
which attempts to use computer vision methods to
distinguish between good and bad papers, large confusing tables were
identified as a key feature of \emph{bad} papers, not an essential
feature of good ones.

So how did computer vision transform from a mathematical discipline
based on geometry to an empirical, quantitative discipline based on
benchmarks? This transition was gradual: we can see its seeds as early as a
debate at ICCV 1999, referenced by P9, between Jitendra Malik and
Olivier Faugeras \cite{triggs2000vision}. In that debate, Malik argued
that computer vision should focus more on probabilistic modeling and
perception, rather than methods based in geometry, while Faugeras
rejected empirical computer vision as unfalsifiable, advocating for the mathematical rigor of geometry.

The publication of Krizhevsky, Sutskever, and Hinton's ``ImageNet
Classification with Deep Convolutional Neural Networks'' in
2012~\cite{NIPS2012_c399862d} marks a turning point for empirical
evaluation. This paper is historically significant for setting off the
deep learning revolution, and its design and writing served as a
foundation for the thousands of deep learning-based computer vision
papers that followed. The paper's central
argument is that several ``new and unusual
features'' lead deep convolutional neural networks to significantly
outperform other methods. These features include
rectified linear units (ReLU), GPU-based training, and regularization
techniques like data augmentation and dropout. Crucially for our
story, however, this argument is made by way of a table, shown in
Figure~\ref{fig:table_examples} (c), with the best performance in
bold. Neural network papers are obligated to use empirical
evaluation, as there are insufficient theoretical guarantees for these models and
they are difficult to evaluate otherwise. Over the following years,
many papers followed Krizhevsky et al.'s lead, showing that deep
convolutional neural networks outperform existing methods on other
central problems like object detection and semantic
segmentation. We can see a corresponding increase in both the average number of tables per paper, as well as the fraction of papers containing at least one table in Figure \ref{fig:table_counts}. While the prevalence of figures has remained relatively constant, tables have become significantly more common. While only 75\% of CVPR 2013 papers had a table, 95\% of CVPR 2021 papers did, and the average number of tables per paper has doubled from two to four. While not all of these tables are the kind of results table we are discussing, the change is striking.

Like teaser images, however, this style of table arises in computer
graphics before entering computer vision; see
Figure~\ref{fig:table_examples} (a) for an early example. Early graphics results tables
primarily showed runtime comparisons, rather than accuracy or quality
evaluations. These tables are used in computer vision for showing machine
learning performance at least as early as 1999 (Figure
\ref{fig:table_examples} (b)). Interestingly, tables from this era
usually had methods in the columns and different data examples in the
rows, in contrast to the later tables which have evaluation metrics in
columns and methods in rows. This swap also aided readability, as it is
easier to scan vertically than horizontally \cite{luckiesh1941effect}.

As competition has heated up, results tables have grown. For example,
compare the table in Figure \ref{fig:table_examples} (c), from a 2012 paper, to
the table in Figure \ref{fig:table_examples} (d), from a 2021 paper. The
benchmark remains ImageNet, though performance has surged from 40\%
top-1 accuracy to over 85\%, but the competing state-of-the-art includes
dozens of models and differ by only
fractions of a percent. Even highly geometric papers, like the example in Figure \ref{fig:table_examples} (f), now involve empirical evaluation.

More broadly, in computer vision there is a widespread assumption that research inherently involves competition between technical methods. Several of our participants described academic publication using free market metaphors. P2 explains his research output: \textit{``publication was really fast during that time because there was not a lot of competition.''} P3 explained that CVPR has become inaccessible to students because \textit{``supply and demand''} have raised the standards for publication. P4 described vision as \textit{``very industrial''} and gave examples of techniques that students use to optimize their arXiv submissions to reach as many eyes as possible. 

With these tables  buried deep within the paper, it may seem like their effect on attention is minimal. However, a competitive benchmark result attracts attention on its own. Once a researcher has the lead, other papers must compare against their benchmark and cite them. P6 compares benchmarks to arcade game leaderboards: \textit{``If it's an established benchmark... there's somebody who has the lead right? Like you would have on like a classic video game arcade machine, right? It's like I have to just get my initials at the top right? That's exactly what they're doing, and that's frustrating.''} P5 and P8 advise their students to avoid working on research which forces them into competition with large companies. P5 explains, \textit{``I tell them, don't work on problems that, you know, a lot of people are working on right now, you can't possibly compete with Facebook, Google, Amazon because you're not as computing heavy.''} While students could still develop innovative ideas for those problems, they likely would not beat the benchmarks that large companies have set without leveraging similar computing power and would not attract attention. Because students will not be able to compete with large companies for attention on these problems, they are not worth studying.
%if you do that you can see that you get all stressed out.''}

In summary, the development of the results table shows the transition
of computer vision from an applied mathematical discipline to a
quantitative-empirical one. As vision started relying on empirical
evaluation, it adopted a style of results table which was used for runtime
benchmarking in computer graphics, and as evaluation benchmarks became
established for vision problems, these tables grew in size and
importance. Today, competition on benchmarks is the organizing
principle of the discipline; new methods must demonstrate that they are
empirically more effective than existing methods to be accepted, and
the results table is an essential part of the paper. This element
showcases several of the same themes as teaser images: influence from
computer graphics, and commodification, in the form of measurable
improvements over prior work. It also echoes patterns observed in HCI
more broadly regarding quantification: once a phenomenon, in this case 
the quality of a method, has been measured, that measurement creates 
and constrains possibilities for action \cite{pine2015politics}.

\subsection{Disseminating the Contribution}

\begin{figure*}
    \centering
    \begin{tabular}{c c c}
    \includegraphics[trim={0 4in 0 0},clip,scale=0.25]{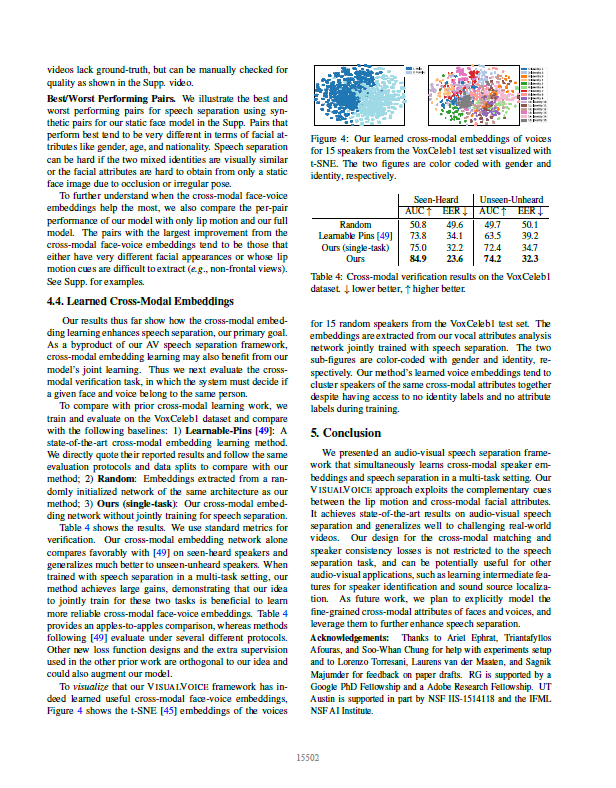} &
    \includegraphics[scale=0.11]{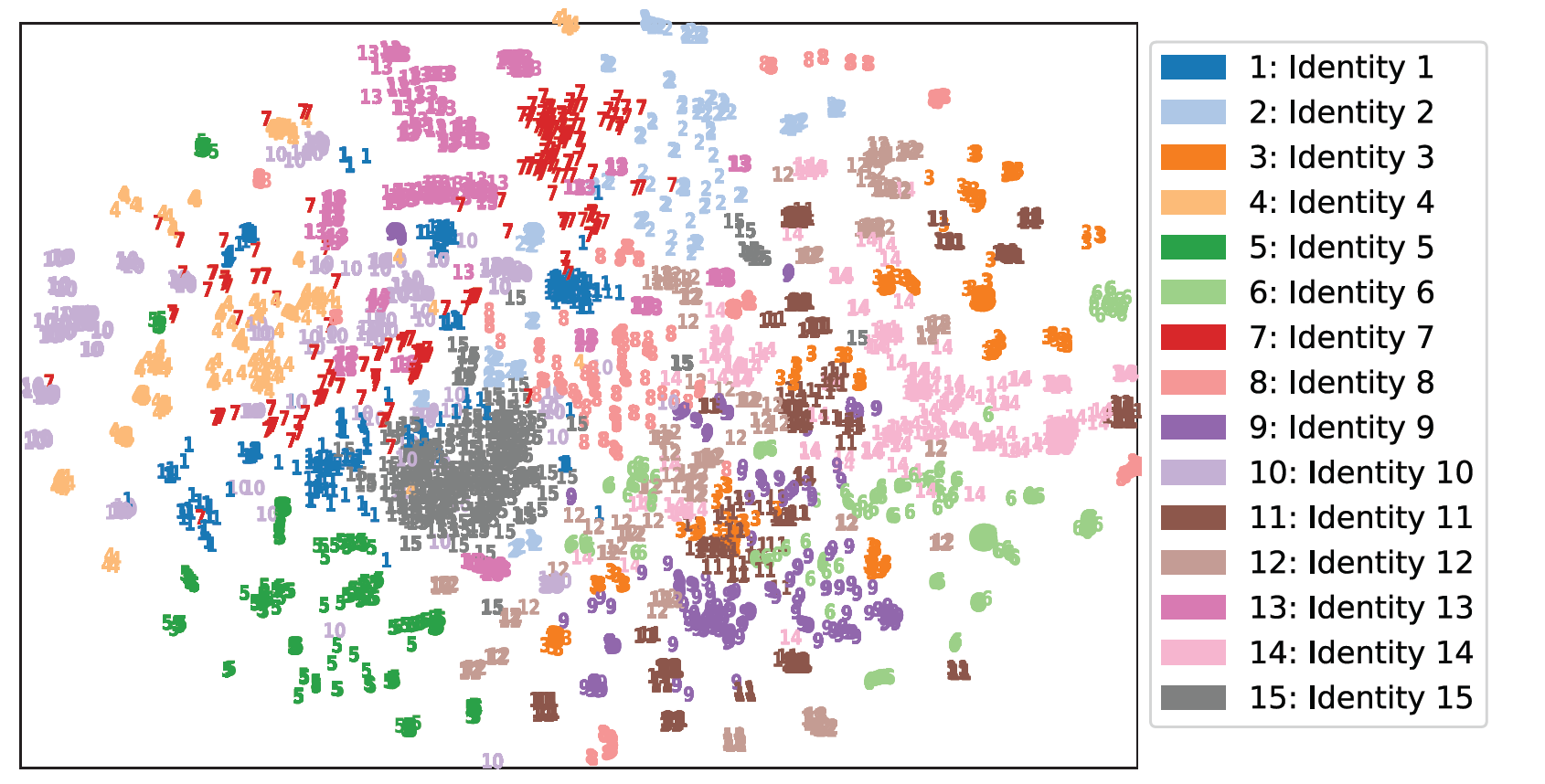} &
    \includegraphics[scale=0.09]{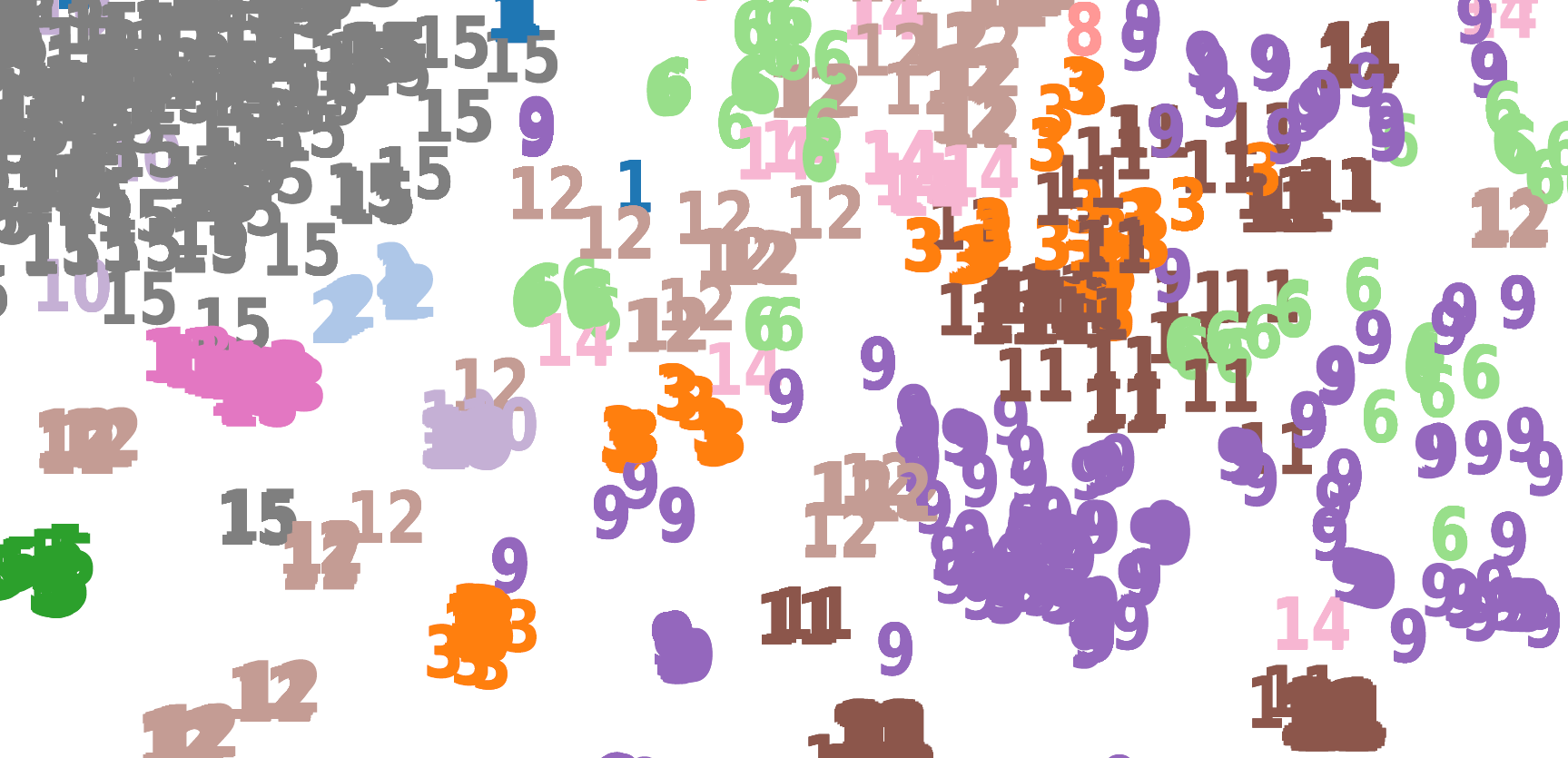} \\
    (a) 25\% scale.&(b) 100\% scale.&(c) 300\% scale.\\
    \includegraphics[scale=0.2]{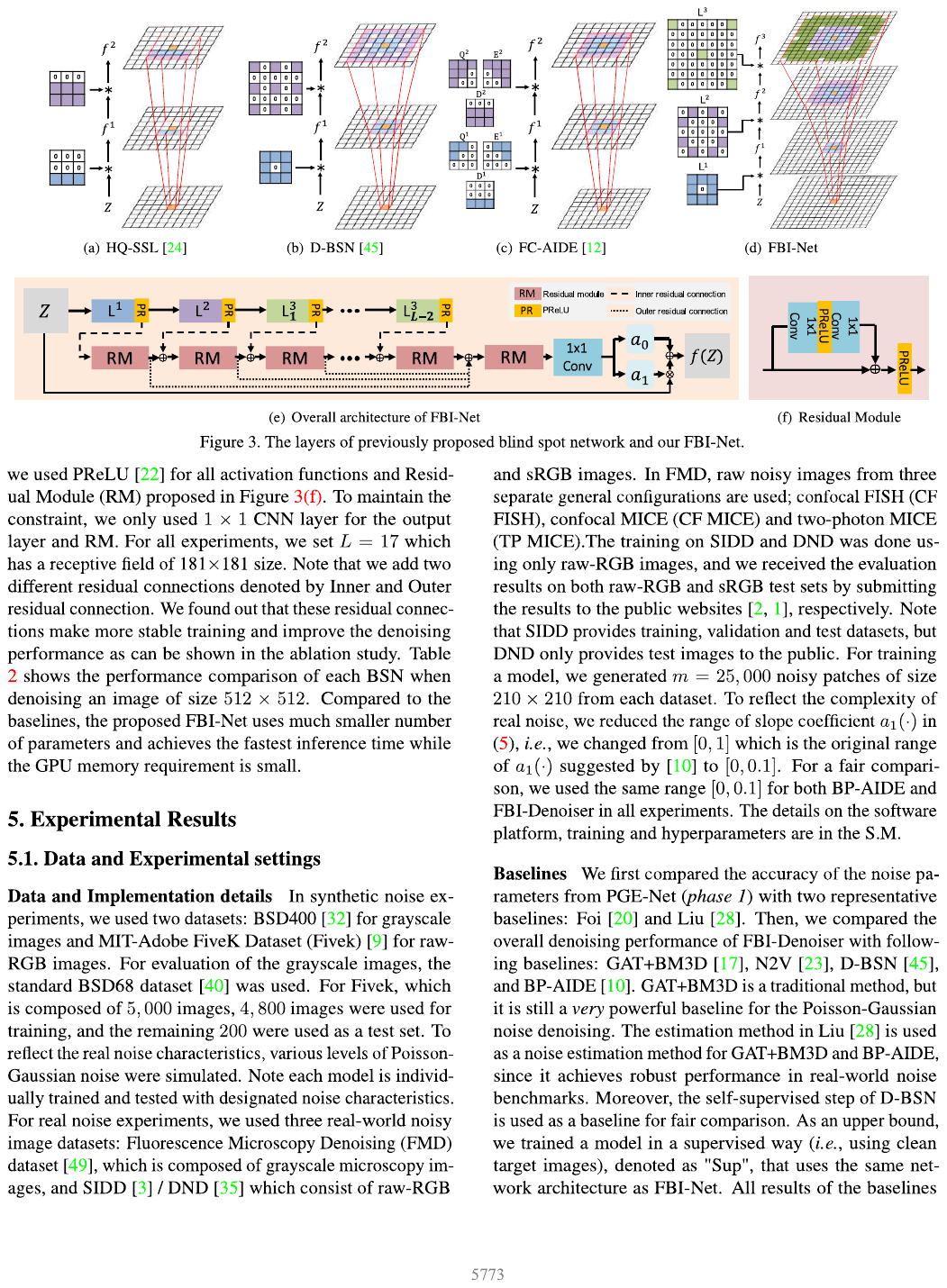} &
    \includegraphics[scale=0.2]{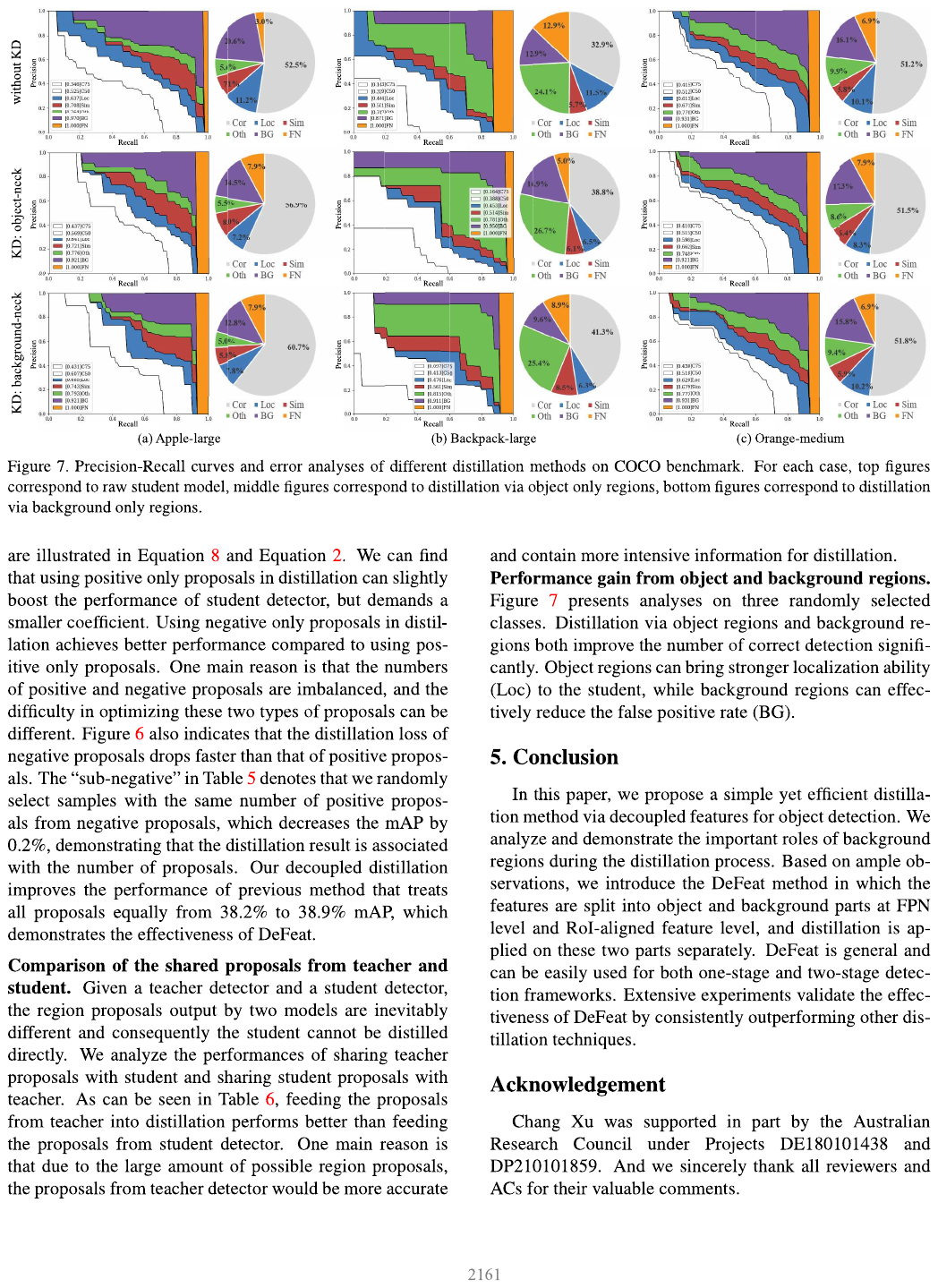} & \includegraphics[scale=0.2]{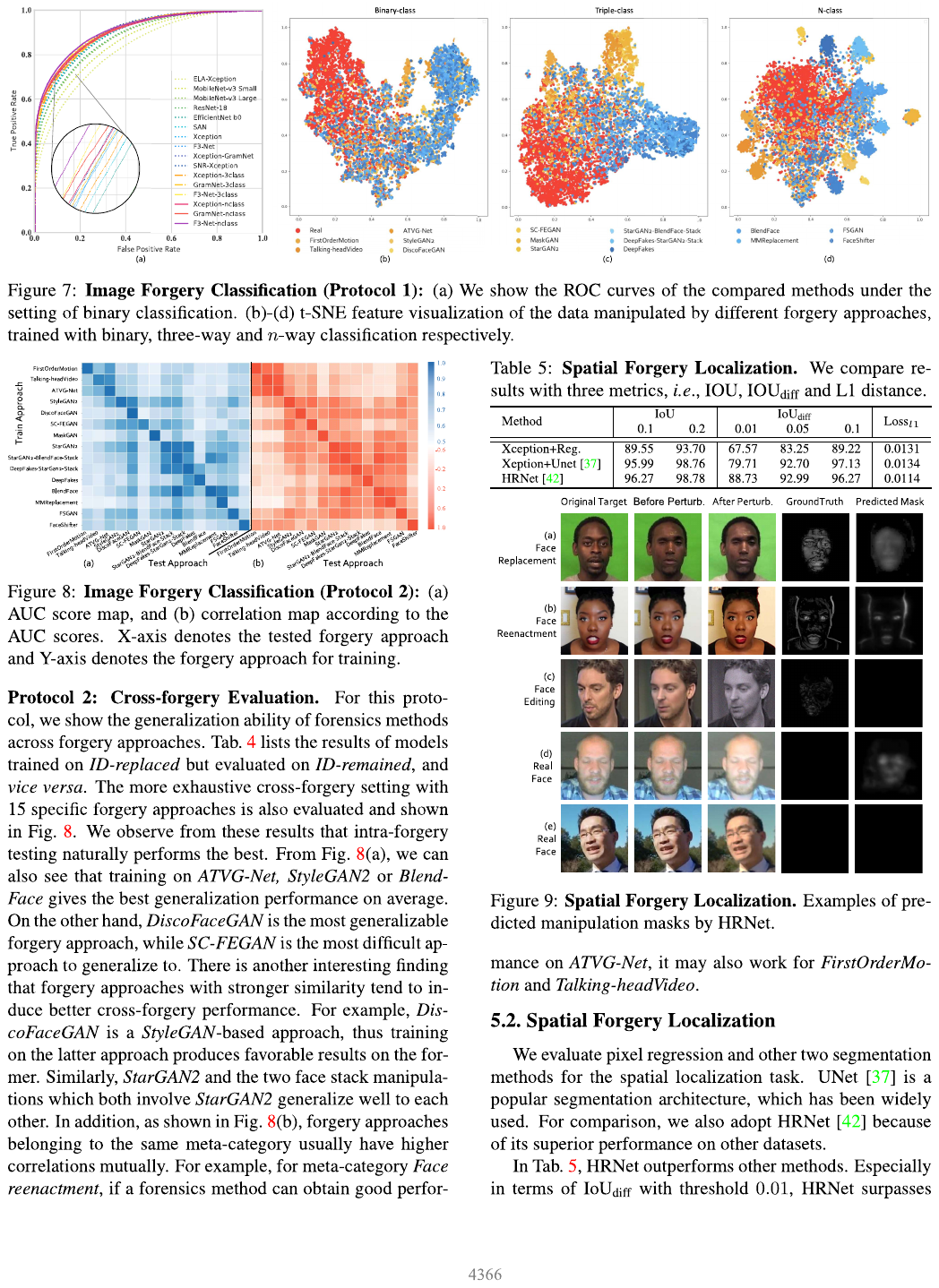}\\
    (d) A page from \cite{byun2021fbi} &(e) A page from \cite{guo2021distilling} &(f) A page from \cite{he2021forgerynet}
\end{tabular}
    \caption{Top: three screenshots at different sizes of a figure from \cite{gao2021visualvoice} which is too small to reproduce using a standard office printer. Bottom: three examples pages from CVPR papers with highly information-dense figures in small page regions, shown at 20\% of original size.}
    \label{fig:nested_figure}
    \Description{Six subfigures: (a) a screenshot of a research paper with two scatterplots in the top right corner, occupying a tiny fraction of the page. (b) and (c) are zoomed versions of the scatterplot, revealing that each data point is represented with a digit drawn on the pdf. (d) shows a page with a dense neural network diagram in the top third with tiny mathematical annotations. (e) shows a page with nine ROC curves and 9 pie charts in the top third. (f) shows a page with three figures and a table in half the page area, one figure contains three scatterplots and a line graph, the second contains two matrix heatmaps with tiny labels for each of dozens of columns, and the third contains 25 images of faces.}
\end{figure*}

\subsubsection{The PDF and digital proceedings}

The final element we explore is the Portable Document Format (PDF)
itself, and the way that authors engage with it as a medium. PDF was a derivative of Adobe's
already-successful Postscript  language for specifying
documents, and a proprietary file format until 2008. In an article
from 1998, Kasdorf compares the PDF format to SGML (Standard
Generalized Markup Language), and argues that both formats should be
used on the web: markup languages for screen-based content and
page-based languages for printed content \cite{kasdorf1998sgml}.

However, in the context of computer vision research, the PDF is now
more of a screen-based format than a printed one. The PDF viewer has
steadily replaced paper as the medium for reading research articles,
which has led researchers to design their articles for viewing on 
screens, rather than as physical, printed research
papers.

Our participants ascribe this shift to an aesthetic factor: color. P9 explains that a huge draw of a conference like SIGGRAPH was
its color proceedings: \textit{``SIGGRAPH was always in color. It was
  absolutely beautiful color. Much more expensive.''} P10 points to a
turning point where researchers started preparing all of their figures
in color: \textit{``There was also a turning moment at some point, I
  think it was around 2004-5-6 something like that. So before that, it
  was black and white plus color images at the end. Like in a set of
  separate full color plates as they call them. It was difficult to
  decide which figures to put in color, since visualizations had to be
  fully redesigned for black and white.''} Once some figures could be
in color, authors had to choose which of their figures to leave in
black and white, which was a difficult choice. P10 explains his
solution: \textit{``the only simple solution I found is you don't
  decide. You just leave it like this and then they complain, well
  this is not visible...[so] you insert the URL of your webpage.''}

But the existence of URLs, and particularly hyperlinked URLs,
points towards another motivating factor for digital proceedings. Before the
web, articles were written, submitted, and published on paper. But
through the 1990s and 2000s, a series of rapid technological changes
took place, transitioning to digital submission and dissemination. The transition to digital proceedings, like other trends, started in
SIGGRAPH, which produced the first fully digital proceedings on CD-ROM
in 1993 \cite{brown2007history}. CVPR distributed digital proceedings
as early as 2005, allowing color figures without the
cost of color printing. P9 explains that in 1999, digital
versions of papers were hard to read: \textit{``most papers were in
  postscript and the postscript viewer was kind of sucky. Didn't
  really look very good on the screen.''} P7 found the transition to PDF freeing: \textit{``I've seen
  figures have become a whole lot more common...the PDFs are now in
  color so you have a whole bunch of other choices.''}

One such choice is the high
resolution figure, which exploits the scalable nature of vector
graphics and the zoom feature of PDF viewers to put more data in a PDF without exceeding the page limit. Several examples of
this trend from CVPR 2021 are shown in Figure \ref{fig:nested_figure}. While these figures are not overly complex
or poorly designed, they are too small to be readable on a 300-dpi
printed page.

Several of our participants mentioned these high resolution
figures. P6 and P10 are frustrated that they cannot print some PDFs
because the figures are too small. P10 ascribes these figures to the
increasing complexity of research: \textit{``the tendency of grouping
  more small figures into a kind of collection... I wonder, is this
  because of the limited page count that people try to avoid the
  whitespace around the figures and then they put everything in one
  big figure because then you can save on the whitespace? Or is it
  simply because really...the techniques that we're describing
  nowadays are so, so much more complicated? So you need them to...put the architecture of the whole thing in detail
  there. Otherwise nobody understands it.''} P7, on the other hand,
ascribes it to digital submissions: \textit{``In '94, you turned in
  your paper by actually having a piece of large paper that was
  probably 24'' wide by 30'' or something...you had
  columns of text from LaTeX and then your figures had to be glued in
  place in what's called `screen,' right? So you actually have to
  make sure that they didn't copy badly.''} When papers were submitted
by mail, the medium of the ``camera-ready'' draft was a significant
limitation on the kinds of visual content that could be included, but now the only limitation is on the number of pages, so authors use smaller figures to include more content. The small size of these figures, especially in comparison with the large front page teaser image, creates a visual hierarchy, highlighting the differences in purpose between these figures. The teaser attracts attention, the dense figures provide details for those who zoom in, and the impression of details for those who do not.

The digital nature of contemporary research papers was a sensitive topic
in our interviews. Half of our participants (P1, P2, P3, P4, P5, and P8)
expressed attachment to and nostalgia for their printed papers, and
were saddened when asked when they stopped reading on paper. For P1,
the time spent at the library made the papers physically significant:
\textit{``[the papers] were all me standing at a Xerox machine with a
  journal...I didn't find it as annoying. I found that it gave me
  something like a kind of a physical connection to the paper. This
  was sort of like, it's mine. I've watched it go by, you know.''} P2
remembers that the SIGGRAPH proceedings \textit{``were like flipped to
  the point where like it's the threads were like bare.''} He's kept
all of the papers he printed during his master's degree and revisits
them from time to time: \textit{``I still go back, you know, look at
  like oh this crap that I thought about these things, right?''}
Finding and photocopying papers made them precious; he thinks there is
\textit{``a kind of a correlation between that sense of, like scarcity
  of it, like how precious the paper was, because it's so hard to find
  versus the amount of care you give it.''} P4 fondly recalls his
advisor taking papers out of a filing cabinet and photocopying them
for him: \textit{``the amazing thing about your advisor is you have
  this filing cabinet of stuff that's already pre-copied, you know,
  every paper he thought was interesting.''} 
  
These sorts of relationships to printed papers are interesting from a design
perspective. They echo the findings of Odom et al.\ that we preserve
designed objects which have functional, symbolic and
material-aesthetic value \cite{odom2009understanding}. When a researcher prints out and writes on a paper, the paper gains symbolic significance, a \textit{``physical connection''} which endures over time. That makes printed pages less disposable than digital ones; there is a more tangible opportunity cost to creating them. From this perspective, digitization is an essential component of commodification, as it separates the research from its paper and thus its role as symbolic object. While reading digitally is significantly more environmentally sustainable than printing, shipping, and photocopying paper \cite{aydemir2020environmental}, many of our participants who primarily read on screens miss the era of physical papers.

In contrast, P7 prefers PDFs because they allow him to use a screenreader: \textit{``I use an app that reads the pdf aloud to me...I can get most of the paper from that, I still have to read equations.''} Despite being an advocate for making computer vision papers more accessible for the community, he explains that conferences are reluctant to put accessible equations in their templates because of its impact on file size: \textit{``basically the conferences keep using templates that---so there's a trade off by adding this stuff to your file, your file gets bigger. And it turns out that the package, that one of the packages is pretty good for this, if you do it badly, it blows files up huge...I tried to get the guys from CVPR to use it and it just didn't happen, right?...I've just lived with the fact that it's not there.''}

The juxtaposition between high resolution figures and inaccessible
equations is ironic. These figures can have
surprisingly large file sizes---the PDF shown in Figure
\ref{fig:nested_figure} (a) is a stunning 36MB, more than 10 times the
size of a typical CVPR PDF. But computer
vision proceedings remain inaccessible, justified based on file size concerns. These figures also point to
the vestigial nature of the page limit. Originally, page limits were
put in place to minimize printing costs. But in the era of online-only
proceedings, there is no financial reason to keep papers page-limited. In fact, the main limiting factor on conferences is the number of reviewers, rather than the lengths of accepted papers. As P8 describes: \textit{``for the last...five years or more, there has been talk every single cycle, oh, you know, we don't have enough qualified reviewers. We have way too many submissions. Everyone's way too overwhelmed to do the reviewing.''} 

\subsubsection{Faster publication through arXiv}

The digital materiality of the PDF affords an alternative
publication process that came up in several of our interviews (P4, P5, P8): the preprint server arXiv. P5 says arXiv is a major source of anxiety for his students: \textit{``So   my students...had the unpleasant experience of, you know, finishing   a paper when they're just about to submit and they saw a paper on arXiv that did almost the same thing. Like, oh my, months of work just went down the drain.''} This emotion, the sense of loss when a
project becomes unpublishable because someone else got there first,
echoes Su and Crandall's observation of ``selective amnesia''
\cite{su2021affective}, except in addition to old papers quickly
becoming obsolete because something else is better, current papers may
become obsolete because they are no longer first.

Conferences originally gained popularity in computer vision because they
allowed research to make it to print faster than journals. According to P3,
\textit{``it used to take almost two years from the time a research is
  done and the PAMI [Pattern Analysis and Machine Intelligence journal] paper
  would appear. And so the conference has started becoming more and
  more important at that time. So that's the origins of ICCV and CVPR
  conferences becoming far more popular than journals. So during the
  1980s and early 90s, the journal was the thing.''} But now, if
research is primarily disseminated by PDFs posted on arXiv, it seems
like the same process is occurring again: conferences behave like
journals and arXiv behaves like a conference. As P8 explains, researchers still submit to conferences for prestige:
\emph{``In order to get promoted in order to get a job, you need that
  stamp of approval. And it's a pretty strong signal to get papers
  accepted to a selective conference...this is really kind of the, you
  know, biggest marker, you know, the biggest benchmark by which you
  are evaluated.''} As arXiv replaces conference proceedings as the fastest communication medium, the conference is left primarily to evaluate papers.

ArXiv changes the writing process as well. As P9 explains, \textit{``I find that the amount of time and effort put into each paper has gone downhill. For my papers, I would put a huge amount of time, especially in the intro...And recently I have been doing less and less of that and basically because the students are saying, ah you know, the new kids they don't even, they just skip the intro, they just go directly to the method. So I feel like, oh my god, nobody's even gonna read my beautiful prose!...But also I think because the field progresses faster, papers become obsolete much quicker. So it might be reasonable not to spend so much effort on a single paper if you know that in a year it will be obsolete.''} In other words, the author now writes differently in order to better fit the faster publication system, spending less time on introductions in order to spend less effort on a paper which might quickly become obsolete.

To summarize, the technologies underlying the publishing process in
computer vision have changed rapidly over the past three
decades. Today, proceedings are published online, and most researchers
read PDFs on a screen, rather than research papers in print. The loss of physical research papers
affects readers' attachment to those documents, as virtual papers
cannot hold significance as domestic objects in the same
way. Meanwhile, authors have taken advantage of this fact and
increased the resolution of their figures to bend conference page
limits, and started publishing on arXiv to quickly attract attention to their new results. Again, we see the same themes: a pattern in computer vision
was preceded by a similar shift in computer graphics, which has
contributed to commodification. In this case, the conference publishing system has shifted from a means of scholarly communication to a means of evaluating research. Acceptance to a computer vision conference serves as a marker,
not just of peer-reviewed technical correctness, but of sufficient
novelty and significance to warrant high scores from reviewers, a
\textit{``stamp of approval''} for employment or promotion. In other words, improvements in writing tools have shifted the burden of evaluating new research work from the authors to the peer reviewers, placing significant labor burdens on the peer review process.

\section{Discussion}

Using a media archaeology approach, we have described the development of
three aspects of the design of the contemporary computer vision research
paper. First, we saw how teaser images, titles with acronyms, and videos advertise the contribution of a paper, and how attention from arXiv and social media has become more important. Second, we saw how the results table was
introduced for measuring the significance of a contribution, and became ubiquitous. Finally, we looked at the transition from paper to PDF as a medium for disseminating research contributions and its new affordances for figure color and density. These trends have a key
commonality: they make research papers easier to consume
visually and more readily disposable.

These shifts showcase the changing material reality of academic publishing. The negligible reproduction
cost of digital documents has reduced the labor required to communicate research. That has shifted the labor burden onto the attention of peer reviewers and online communities of scholars. The peer review process is now governed by self-imposed
limiting factors, born of a desire by conferences to signal prestige
through low acceptance rates. While we observe these changes in computer vision, they may soon appear in other disciplines as well.

When we treat online attention from peers as a kind of labor---something which allows authors to generate value from the research activities of their peers---we see a clear explanation for our visual trends. Authors who write more papers which are
easy to understand at a glance, easy to promote on social media, more 
obviously novel, and more significant in the eyes of reviewers attract more attention. Attention confers a variety of benefits including citations, interest from future mentors, students, employers, and employees, and increased progress on specific research problems. These benefits increase the value of the author's research.

Authors seeking attention can develop design innovations, like more readable tables, but the attention economy ultimately privileges large industry 
labs. Discussion around the CVPR social media ban echoes these attempts to prevent competition between large industry labs and graduate students. But even social media bans cannot prevent the design of the paper itself from attracting attention. In reality, the value of that attention is great enough to motivate industry labs to invest in publishing. In addition to the marketing benefits, those companies can leverage scholarly attention to signal 
~\cite{FeldmanMarch:1981} the innovation and pedigree needed to recruit the best computer vision researchers.

We believe this pattern exemplifies a broader trend of exploiting the attention of researchers. While authors derive prestige and career progress from online attention, the researchers who read those papers do not. Instead, paying attention to the state of the discipline is essential for doing career-progressing research of their own. While staying current has always been a part of academic work, the growth of the discipline puts pressure on readers, increasingly treating them like a crowd of interchangeable peers whose attention is needed to make new research prestigious. This process resembles proletarianization\footnote{In Marxist theory, the process by which people are assimilated into the working class, causing alienation.} \cite[Ch.\ 3]{buenoattention}, and may cause the ``malaise'' observed by Su and Crandall \cite{su2021affective}. Just as these trends in computer vision were foreshadowed by trends in computer graphics, similar trends may follow in other fields. In a recent paper examining paper titles and abstracts, Ken Hyland argues that an attention economy has begun to form widely throughout academic publishing \cite{hyland2023academic}. We encourage future study of the attention economy to focus on both the labor performed by peers, as well as the sort of visual evidence we have discussed.

These problems become more egregious when considered alongside the much broader labor issues involving data labeling in computer vision \cite{paullada2021data,scheuerman2021datasets,denton2021genealogy}. Workers have their intelligence commodified and exploited by researchers who are themselves responding to the commodification of their work. In this way, commodifying the labor of relatively privileged academics can exacerbate widespread exploitation of a less privileged workforce.

These problems also echo recent discussions regarding generative artificial intelligence. These technologies (some of which originate from CVPR, e.g.\ \cite{li2023q}) show risk of producing online disinformation \cite{vykopal2023disinformation}, displacing artists and writers \cite{jiang2023ai}, and creating problems for future curators and historians regarding provenance and authenticity of historical images \cite{hazan2023dance}. For example, as Jiang et al.~discuss, the science fiction magazine Clarkesworld was forced to pause submissions due to excessive AI-generated submissions after ChatGPT went public in 2023 \cite{jiang2023ai}. While authors submitting to computer vision conferences are not yet producing their papers using AI text generators, there are parallels regarding a deluge of content that places the burden of evaluation on the attention of reviewers. While bans on AI-generated content may slow the growth in the short-term, they do not resolve the fundamental commodification of attention, and may fail as new tools continue to blur the boundary between human and AI-generated writing.

The main implication is that \textit{ad hoc} responses to symptoms of the attention economy, like banning discussion on social media or punishing low effort peer reviews, will not change the pressures which are at play. Instead, computer vision is faced with a wicked problem \cite{buchanan1992wicked}, spanning peer review, hiring, promotion, and tenure, in addition to academic discourse, which demands a larger-scale rethinking of the inherent tensions between facilitating scholarly communication and evaluating scholarship. Removing barriers to research communication appears to speed up the pace of research, but also changes disciplinary practices by forcing scholars into an arms race, competing for the attention of their peers.

Our inquiry also carries implications both for the design of the
technologies which support publication and literature reviewing. As conferences
continue to move beyond compatibility with paper, there is growing
need for a LaTeX-compatible, digital-first document file format which
gives authors control of the look and feel of their publication, but
is machine-readable and supports accessibility tools. There is also a growing need for tools to help scholars collect and curate relevant literature. But rather than make this process as fast and frictionless as possible, we encourage work grounded in ``slow design'' \cite{odom2012slow} and ``slow science'' \cite{stengers2018another} which attempts to recreate the preciousness, uniqueness, and care that our participants ascribed to the hardcopy papers they photocopied by hand. We also encourage, in line with the view of attention as labor, top conferences to consider compensating peer reviewers for the essential labor they perform.

Our work carries significant limitations. Our visual analysis is not systematic, and is likely biased towards visibly
obvious trends, as well as trends which are present in published
papers that have been digitized, ignoring the visual culture of
posters, presentations, and rejected papers. As with any study of
recent history, we cannot take a fully objective approach, as our personal
experiences will skew our judgment. Our
interview participants also only represent the views of senior
researchers.  A study of current students, junior faculty, and other
younger authors and the ``tricks'' and ``hacks'' they use in their
papers would be excellent future work.

Additionally, we have only scratched the surface of the media
archaeology of the research paper, and there are numerous opportunities for future work. Alongside the patterns we analyze, sophisticated visual
languages for system diagrams and renderings of visual features have developed. We encourage researchers to study these visual languages more systematically. We have
also neglected to discuss the relationship between these
sorts of readings of research papers with the much larger field of
information science and bibliometrics \cite{ball2020bibliometrics}. A
more quantitative study which measures the relationships between
visual features of research papers and the structures of citation
graphs, or a modeling study like Weng et al.\ conduct on internet memes \cite{weng2012competition}, may prove fruitful.

We worry that the stylistic conventions of the field may be constraining the types of ideas students and faculty are willing to pursue. While this is true for any set of disciplinary norms, the culture of benchmarks in computer vision fosters a particular mindset of technological determinism, where research becomes a matter of either finding the best performing model for an existing task or proposing a new task and constructing a benchmark dataset for it. Within this mindset, it seems inevitable that all possible visual perception tasks have computational solutions, and research is only a matter of finding them first.

We also worry that these stylistic conventions may be contributing to
the safety and injustice issues which currently surround machine
learning \cite{buolamwini2018gender,mehrabi2021survey}. If authors are under pressure to publish more and faster, it
is easiest to do that by overstating the significance of completed work. There are a range of behaviors which contribute to this problem:
from neglecting to explore the limitations of a method, to only
showing favorable evaluations, to outright  fabrication of results. While we do not claim that authors are engaging in such behaviors, we do observe a tendency to write papers like
advertisements and only minimally discuss the downsides of advertised methods. These behaviors become problematic when engineers,
stakeholders, and researchers in other fields who are unfamiliar with 
the reality of conference
publishing may take the claims in papers at face value, and
put minimally tested systems into production based on a general trust
of computation and data.

\section{Conclusion}

In this paper, we have examined the commodification of researcher attention through the visual evidence in computer vision research papers. These changes lead authors to include
teaser images, and the presence of teasers leads researchers to
develop methods which can more easily be presented as products,
ready-made for readers to download, use, and cite. Second, the results
table allows researchers to demonstrate the effectiveness of their
neural network methods, but then the expectation of comparison to
other methods forces other kinds of papers into the same evaluation
paradigm. Finally, desire for color figures led conferences to adopt
digital proceedings, which has led authors to create content which
cannot be viewed on paper, necessitating screen-based reading. Through Bueno, we interpret attention as a necessary form of labor in academic publishing. These trends shift that labor onto the attention of researchers and especially peer reviewers. While better tools to support paper-reading (e.g.\  \cite{qian2019beyond,chang2023citesee,head2022math}) can help streamline this process and reduce the burden on individual scholars, the problem will persist unless scholarly communities address the commodification of scholarly attention. Generally, we recommend that conference organizers approach this situation as a systemic problem, and investigate measures to avoid competition between authors, slow the submission and peer review process, and search for more equitable and efficient ways of distributing attention and prestige.

\begin{acks}
This work was supported in part by a National Science Foundation Graduate Research Fellowship. The views expressed in this paper are those of the authors and do not necessarily reflect the views of the NSF or any other agencies.
\end{acks}

\bibliographystyle{ACM-Reference-Format}
\bibliography{sample-base}

\end{document}